\documentclass[letterpaper, 11pt]{article} 

\pdfoutput=1
\usepackage{times,abstract}
\usepackage[authoryear]{natbib}
\usepackage[pdftex]{color,graphicx} 
\usepackage{amsxtra}     
\usepackage{amsthm,gensymb}
\usepackage{amssymb}
\usepackage{amsfonts}
\usepackage{upgreek}
\usepackage{subcaption}
\usepackage{array} 
\usepackage{lscape}
\usepackage{rotating} 
\usepackage[table]{xcolor}
\usepackage{cellspace}
\usepackage{amsthm} 
\usepackage{verbatim}  
\usepackage{graphicx} 
\usepackage{amsmath} 
\usepackage{amssymb} 
\usepackage{float}
\usepackage{multirow}
\usepackage{array} 
\usepackage{ragged2e}
\usepackage{picinpar} 
\usepackage{rotating} 
\usepackage{fancybox} 
\usepackage{wrapfig} 
\usepackage[all,2cell]{xy} 
\usepackage{epigraph}
\usepackage{grffile}
\usepackage{soul}
\usepackage{dcolumn,ctable}
\usepackage[bf]{titlesec}
\usepackage{setspace}    
\UseAllTwocells
\theoremstyle{plain}

\theoremstyle{remark}

\theoremstyle{definition}

\renewcommand{\i}[1]{\textit{#1}}
\renewcommand{\b}[1]{\textbf{#1}}
\definecolor{tableShade}{HTML}{F1F1F1}

\titleformat{\section}{\large\bfseries}{\thesection.}{1mm}{}
\titleformat{\subsection}{\normalsize\bfseries}{\thesubsection}{1mm}{}
\newcolumntype{V}{>{\centering\arraybackslash} m{.4\linewidth} }
\newcolumntype{d}[1]{D{.}{.}{#1}}   

\numberwithin{equation}{section}

\makeatletter
\let\@fnsymbol\@arabic
\makeatother

\newcommand{\affiliation}[2]{%
   \footnote{#2}
    \newcounter{#1}
    \setcounter{#1}{\value{footnote}}%
}
\newcommand{\sameaffiliation}[1]{%
    \footnotemark[\value{#1}]%
} 

\newlength{\imgwidth}

\begin{document}

\bibliographystyle{plainnat}    
\bibpunct{(}{)}{,}{a}{}{;}

\title{Benchmarking Measures of Network Influence}
\author{Aaron Bramson\affiliation{Riken}{Laboratory for Symbolic Cognitive Development, Riken Brain Science Institute} \affiliation{Gent-Econ}{Department of General Economics, Ghent University} \affiliation{UNCC}{Department of Software and Information Systems, University of North Carolina at Charlotte} %
\and Benjamin Vandermarliere \sameaffiliation{Gent-Econ} \hspace{0.2mm} \affiliation{Gent-Physics}{Department of Physics and Astronomy, Ghent University}%
}

\singlespace

  \maketitle             
  \begin{onecolabstract} 
  Identifying key agents for the transmission of diseases (ideas, technology, etc.) across social networks has predominantly relied on measures of centrality on a static base network or a temporally flattened graph of agent interactions.  Various measures have been proposed as the best trackers of influence, such as degree centrality, betweenness, and $k$-shell, depending on the structure of the connectivity.  We consider SIR and SIS propagation dynamics on a temporally-extruded network of observed interactions and measure the conditional marginal spread as the change in the magnitude of the infection given the removal of each agent at each time: its temporal knockout (TKO) score.  We argue that the exhaustive approach of the TKO score makes it an effective benchmark measure for evaluating the accuracy of other, often more practical, measures of influence.  We find that none of the common network measures applied to the induced flat graphs are accurate predictors of network propagation influence on the systems studied; however, temporal networks and the TKO measure provide the requisite targets for the hunt for effective predictive measures.

\vspace{7mm}
\noindent\b{keywords:} networks, temporal webs, centrality, influence, immunization
  \end{onecolabstract}

\section{Introduction}\label{Introduction}

There are two main strategies to identifying the key agents for disease/idea spread: (1) the discovery of ``super-spreaders'' \citep{kempe2005influential,wang2010community,kimura2010extracting,chen2012identifying,saito2012efficient} and (2) finding effective immunization/removal targets \citep{chen2008finding,yu2010finding, kuhlman2010finding}.  The difference is not the goal of the analysis; both approaches seek to ascertain the actual or potential \i{influence} of each node on the propagation of a property across the network by performing an isolated contingency analysis.  The first approach is some version of variably seeding an infection and determining how well it spreads in each setup \citep{newman2002spread,newman2002email,Dekker2013}. The second approach is some version of setting nodes as firewalls and measuring changes in how the property/idea/disease spreads with different firewalls \citep{chen2008finding}.  By toggling the status of any one node and examining the differences it generates one can ask, ``How much of the propagation is this node responsible for?''  Here we propose a knockout sensitivity analysis on temporally extruded networks that combines the spreading and removal approaches for use as a benchmark test for evaluating the ability of network measures to capture or predict system influence.

The dominant technique to assess individual influence is to take a set of agents and a network of potential interactions among them and simulate the propagation of a property using a variation of SI/SIR/SIS dynamics across the network to see how far and how fast it spreads.  There are variations in the (generated or empirical) network structure used, the number and placement of initial infections, the disease parameters, and with these there are variations in the identified best measure of influence (see \cite{danon2011networks} for an extensive review on the possible variations). The most important lesson from these analyses is that different structures make different targets more effective for immunization.  For example, connectivity on some network structures is resilient to random node removals but sensitive to targeted removal of nodes with certain properties, such as high degree agents in scale-free networks \citep{Albert2000,Callaway2000,Pastor-Satorras2002}. For other network structures, high degree is not the best measure of importance; betweenness, $k$-core, and other measures have been proposed as capturing key individuals in certain specific network structures and real-world datasets \citep{Kitsak2010}.  


In order to evaluate network measures' ability to track influence one must have an independent assessment of that influence -- the ground truth to be matched.  A common way to measure this is to seed the initial infection at each node and measure the resulting spread, typically as cumulative cases for SIR. However, an individual's impact on the dynamics of propagation on complex networks is more nuanced than these simple propagation measures indicate.  Even when a disease starts at node $x$, some later-infected node $y$ may be more responsible for the scope of the spread. In actual disease propagation dynamics \citep{hufnagel2004forecast,Brockmann2013} it is also possible that an agent being infected early \i{reduces} the eventual scope of the infection by altering the set of individuals that agent comes in contact with while infected.  

In light of these possibilities it is clear that one must analyze how the full dynamics unfold in order to correctly assess influence over those dynamics.  To incorporate the temporal aspect into our influence analysis we capture the infection propagation in a temporally extruded network structure called a ``temporal web'' -- a variant of temporal networks \citep{Holme2012,holme2015modern} in which the interactions extend across time creating a single acyclic digraph rather than layered networks \citep{bramson2015dynamical,michail2015introduction,speidel2015community}.  This temporal web provides a time-extruded version of cumulative cases that we call ``magnitude'' combining both the number of infected individuals and the length of their infections \citep{saito2012efficient}.

To perform the isolated contingency analysis we propose a measure called ``temporal knockout'' (TKO) that combines the super-spreader and immunization approaches and also includes the timing of infections to more accurately measure each agent's influence/impact on the propagation.  TKO is not an alternative network measure for approximating influence, but rather an all-things-considered empirical measurement of each agent's time-dependent potential to change propagation outcomes for use as a benchmark in evaluating network measures. 

First we explain the temporal web construction in more detail, then we describe the process to calculate the disease magnitude and temporal knockout score.  Because the temporal knockout score calculation is computationally expensive, it is desirable to have a simpler proxy measure, or set of proxy measures, that accurately reflects agent influence.  Toward this end we run a battery of experiments on small world and scale-free networks and evaluate the effectiveness of some standard flat/static network measures to capture influence using the TKO scores as a benchmark measure.  Although the limited evaluation of network measures presented here is indicative of the need for improved ways to capture propagation influence, our focus here is the presentation of TKO as a standardized benchmark metric for performing such investigations.


\section{Approach}

Our analysis proceeds through the following steps: (1) create collections of scale-free and small world base networks; (2) build temporal webs encapsulating a fixed set of potential interactions for each one; (3) simulate propagation dynamics across each temporal web for each agent of each network; (4) calculate the temporal knockout of each node in the temporal web; (5) generate the flattened network and analyze the flat networks using centrality measures; (6) examine the degree to which the flat network measures capture the agents influence as measured by TKO.

\subsection{Network and Disease Parameters}
We simulate the spread of an infectious disease using an agent-based model realizing SIR and SIS dynamics.  Our networks have 200 agents connected in either a small world or scale free network with $800$ and $784$ edges respectively.\footnote{The small world base networks are undirected connected Watts-Strogatz networks where each agent is connected to $k=8$ neighbors and the probability of rewiring is set to $p=0.025$. The scale-free base networks are undirected Barabasi-Albert networks with $m=4$ as the number of edges to attach from a new node to an existing one. The networks were generated using the implementation of the python package NetworkX \cite{Hagberg2008}.} For each combination of network type and interaction probability - 0.10, 0.15, and 0.20 -, we generate $25$ instantiations ($150$ total). We note that the SIR and SIS versions of a given combination run on the same instantiations, thus using the same link activations at each time step\footnote{In each iteration of the model, the probability that a given link is activated is
\[
p_{ij} = \frac{k_{j}^{-1}}{\sum_{K_i}k_{n}^{-1}}
\]
with $k_j$ being the undirected degree of agent $j$, and the summation in the denominator is over each network neighbor ($n$) of node $i$ (written $K_i$).}. There is one initially-infected agent per run and we perform a run of the model using each agent as the initial agent for each of the 25 instantiations of each network type. Each infectious agent has a probability to infect susceptible network neighbors and as already mentioned we run the full battery of simulations using probabilities 0.10, 0.15, and 0.20.  In each period, the probability of infectious agents converting to recovered/susceptible (I$\rightarrow$R and I$\rightarrow$S for SIR and SIS models respectively) is $1/15$.  Each run lasts 200 periods; this is typically sufficient for SIR dynamics to run their course, and is used for SIS models for parsimony of analysis.

\subsection{Building a Temporal Web}

\begin{figure*}[ht]
\centering
\includegraphics[trim={0 5cm 0 5cm},clip,width=0.92\textwidth]{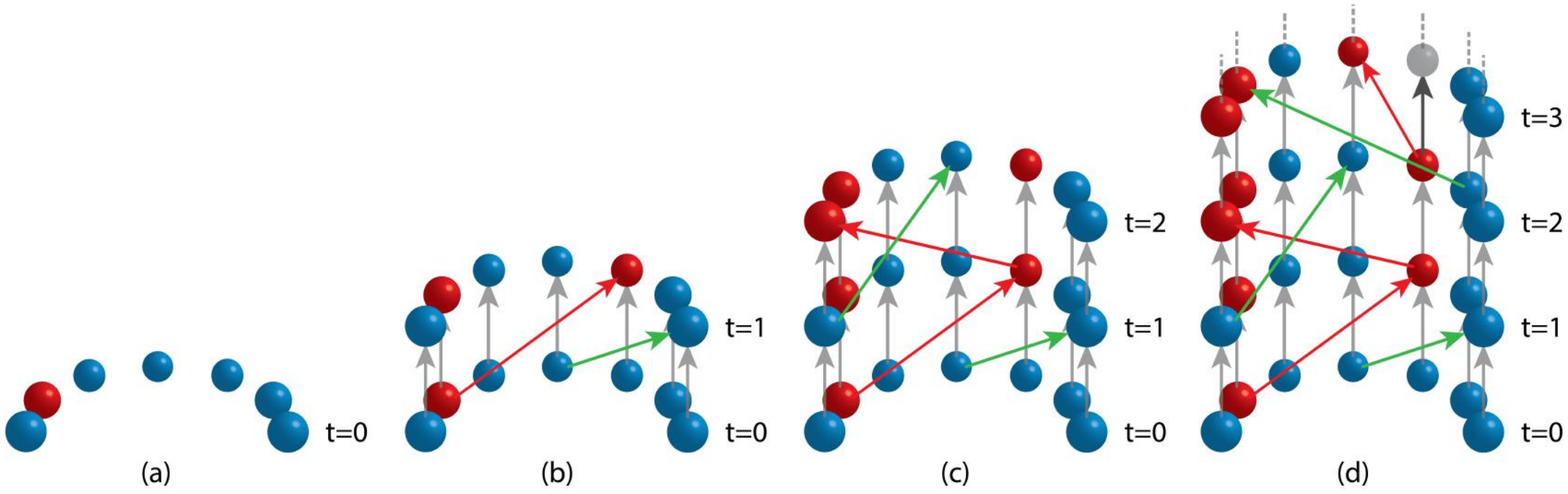}
\caption{A simplified example of building a ``temporal web'' style intertemporal network from state-change and interaction data for an SIR model. This procedure differs from temporally layered networks in that the interaction edges are cross-temporal to capture simultaneous updating in the generated data, thus creating a single acyclic directed graph across time.  }
\label{building}
\end{figure*}

We run our simulations using simultaneous updating so that each agents' state at $t+1$ depends on their state at $t$ and interactions initiated at $t$.  When represented as an intertemporal network the interaction edges therefore run across time from agents at $t$ to other agents at $t+1$ in addition to ``inheritance edges'' from each agent at $t$ to its $t+1$ self (see Figure \ref{building}).  We call this version of intertemporal networks a ``temporal web'' because it produces a single acyclic directed graph across time rather than connected layers.

We first build the temporal web ``skeleton'' that includes all of the state changing and interaction probabilities which may be needed for any particular run.  With non-adaptive interaction probabilities, who interacts with whom \i{and when} all become fixed for those runs.  Therefore when we run the simulation using each agent as the initially infected agent, the overall dynamics are kept constant while we monitor the propagation so that the only difference is the initial agent.

\subsection{Disease Magnitude}\label{DiseaseMagnitude}
The temporal structure facilitates a variety of new measures, which are defined and explored elsewhere \citep{Nicosia2013,Pfitzner2013,bramson2015dynamical}. Specifically for epidemiology it becomes natural to switch to a temporally extended refinement of the standard cumulative cases measure.  Rather than (or in addition to) reporting the number of agents that are ever infected, the disease \i{magnitude} is calculated as the number of agent-times (i.e., nodes in the temporal web) that are in the infectious (or exposed) state.  It is equivalent to the cumulative sum of the number of infectious agents across iterations \citep{saito2012efficient}.  This measure better captures disease morbidity because it accounts for both the number of infections and how long the infections persist -- a large number of very short infections could be considered preferable to a few persistent long-term infections.  Depending on the application, the node count or a normalized version may be preferable -- the number of nodes is the same for all of our experiments described below, so we use the ``raw magnitude.''

\subsection{Calculating the Temporal Knockout Scores}

Temporal knockout (TKO) measures influence by aggregating two levels of contingency.  First we select an agent from the population to be initially infected and run the disease model while capturing each agent's state and interactions at each iteration in a temporal web.  The resulting collection of infectious nodes (agent-times) embodies the magnitude of the illness contingent on that agent being the initially infected one.  Then for each node in the temporal web we perform a knockout analysis: remove that node and run \i{the same infection dynamics} and measure the difference in the disease magnitude.  Thus for each node we capture the change in disease magnitude contingent upon that agent being removed at that time, contingent upon that particular initially infected agent.  

The initially infected agent at the $t_0$ iteration will have a marginal infection effect equaling the whole magnitude.  Note that removing a noninfectious node at $t_0$ still prevents it from being infected later, which affects the marginal infection score of that agent at $t_0$; however, the pre-infection time nodes for an agent will have the same TKO as the first infected time node; thus the calculation can be performed on just the infected subset and backtracked to earlier times.  Perhaps counter-intuitively this effect can be negative; i.e., it is possible to remove an agent from the system at a particular time and have the overall disease spread \i{increase}.  This can happen when agents that are infected by the knocked out agent would normally have quickly lead to dead ends, but when instead infected later by other agents they spread the disease to many more others.

We perform this knockout analysis for every node in the temporal web to get the marginal infection score conditional on that initial agent.  We repeat this process using each of the agents as the initially infected agent and set each node's TKO score as the average marginal infection score across those runs.  Thus we have the conditional marginal infection spread for each agent at each time step for all possible single-agent disease carrier initial conditions.  This algorithm therefore captures the \i{potential} for each agent at each period to influence the spread of the disease.  

Because TKO is an overt counting of infected agent-times given the contingent hypothetical-empirical results instead of a summary measure we believe that it stands as a reliable benchmark for the influence of each agent (in networked epidemiological systems).    
Also note that TKO's hypothetical-empirical approach means that the change in total infection after a knockout of agent $A_i$ at any time $t_\tau$ cannot be calculated except through the resimulation of the infection dynamics across the rest of the temporal web.  Because of this TKO is thoroughly descriptive, but it is not predictive.



\subsection{Base and Flattened Graphs}

In order to predict which agents are most likely to facilitate diffusion, we wish to compare the TKO identification with measures on flat, non-temporal networks.  Specifically we would like to know how well each of various centrality measures does in capturing each agent's network influence as benchmarked by TKO.  Two versions of flat graphs are relevant here: (1) the base potential interaction network from which the actual interactions were probabilistically generated and (2) the flattened empirically observed interactions. Our results for the base network and weighted and unweighted flattened networks are nearly identical, so we focus on the base network here and leave the flattened network for the supplementary materials.  We have twenty-five distinct base networks for each scenario (although each SIR and SIS pair use the same networks) and for every node in each one we calculate the following centrality measures: $k$-core, degree, closeness, betweenness, eigenvector, and Katz centralities.

\section{Results}

The infection dynamics in our model match other models with similar network structures and disease parameters \citep{Rahmandad2008,danon2011networks}.  We briefly summarize the contagion results in order to provide context for the centrality measures and to facilitate comparisons to other models.  For our SIR models the cumulative cases and magnitude measures are nearly perfectly correlated (0.99458) because the fixed 1/15 probability of I$\rightarrow$R transitions implies a uniform \i{expected/average} infection duration time of 15 iterations.  For SIS models reinfection can multiply an agent's contribution to magnitude but still only be counted once by the number of cumulative cases, so the correlation is reduced (0.93595), but is still high due to the relatively short time horizon for our SIS simulations (200-iterations).
 
\begin{figure*}[ht]
\centering
\includegraphics[width=0.83\textwidth]{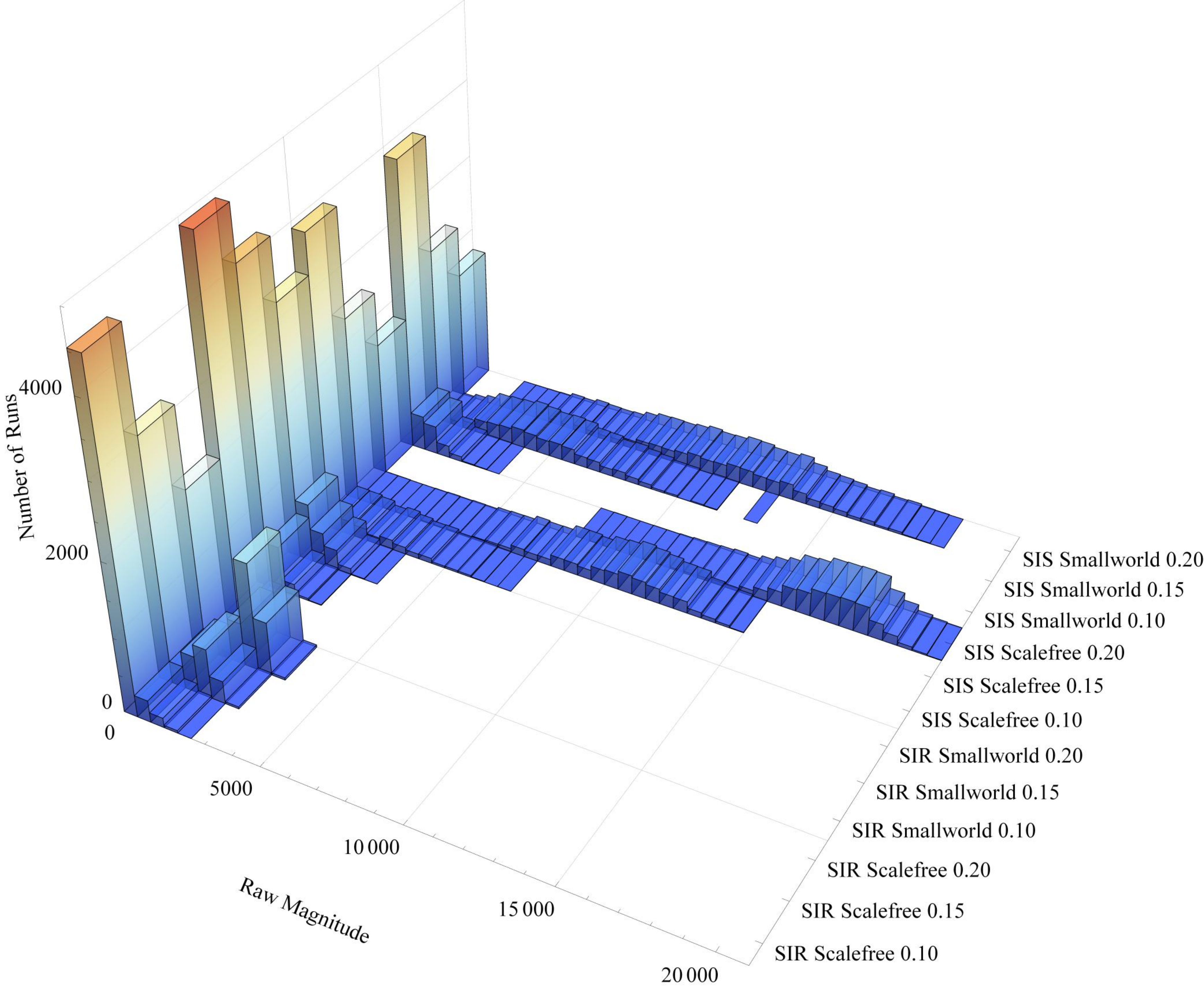}
\caption{Results histogram of infection spread in terms of the number of temporal nodes infected (raw magnitude) across 5000 runs for each scenario (one run initialized at each of 200 agents for each of the 25 base network implementations).  Notice that a very large proportion of runs are ``duds'' in which the infection fails to spread beyond 50 temporal nodes.  The SIS models naturally have greater magnitude values due to reinfection.  These dynamics are typical of SIR and SIS models with similar parameters.}
\label{MagnitudeByCasesHistogram}
\end{figure*}

As seen in Table \ref{MagnitudeSummary} both network types show high variation in magnitude depending on the initial agent; however, when aggregated across the 25 implementations of each network type they reveal similar magnitude profiles (see supplementary material for details).  For ease of reading we present the raw (non-normalized) magnitude scores (i.e., the number of infectious nodes in the temporal web).  As you can see in figure \ref{MagnitudeByCasesHistogram} there are a large number of runs in which the disease never catches on (what we call ``duds'') and although these outcomes drag the mean magnitude down and raise the variance, for our purposes there is no benefit in separating out the duds and, for example, testing the remaining infections for matches to known distributions.


\begin{center} \begin{table}[!ht] \centering
\scalebox{ 0.78 }{
\rowcolors{1}{white}{tableShade}
\begin{tabular}{c c c d{5.5} d{5.5} d{5.5}}\toprule
Infection Type & Network Type & Infection Probability  &  \multicolumn{1}{c}{Mean Magnitude}  &   \multicolumn{1}{c}{Magnitude StDev} &  \multicolumn{1}{c}{Percent Duds}\\  \hline
SIR & Scale Free &  0.10 & 143.352 & 288.549 & 0.625 \\
SIR & Scale Free &  0.15 & 584.744 & 774.628 & 0.482 \\
SIR & Scale Free &  0.20 & 1296.44 & 1142.24 & 0.380 \\ \hline
SIR & Small World &  0.10 & 88.9266 & 131.743 & 0.584 \\
SIR & Small World &  0.15 & 227.321 & 324.207 & 0.457 \\
SIR & Small World &  0.20 & 445.033 & 559.017 & 0.352 \\ \hline
SIS & Scale Free &  0.10 & 548.746 & 1155.19 & 0.593 \\
SIS & Scale Free &  0.15 & 5003.03 & 5237.44 & 0.445 \\
SIS & Scale Free &  0.20 & 10800.6 & 8150.76 & 0.344 \\ \hline
SIS & Small World &  0.10 & 308.734 & 536.106 & 0.557 \\
SIS & Small World &  0.15 & 2526.97 & 2839.26 & 0.433 \\
SIS & Small World &  0.20 & 7036.79 & 5623.18 & 0.333 \\
\specialrule{.08em}{.00em}{.00em} 
\end{tabular} }
\caption{Results summary of infection spread for each model variation.  Each row aggregates 5000 runs (one run initialized at each of 200 agents for each of the 25 base network implementations). Duds are defined as runs in which the raw magnitude is fewer than 50 agent-times. } \label{MagnitudeSummary}  
\end{table} \end{center} \vspace{-10mm}

\subsection{TKO vs Magnitude Correlations Results}
We first compare the TKO score of each agent to the initial-agent resulting magnitude in order to evaluate whether this standard measure of influence effectively captures a node's ability to spread disease.  The TKO algorithm accounts for the idiosyncrasies of the agent interactions across time, but as a result it assigns scores across time as well.  In order to compare TKO node scores to initial-agent-spread scores we first need to aggregate them to the individual agents. 

For each node we determine two versions of TKO: (1) the proportional change in the number of infectious nodes and (2) the change in the fraction of nodes that become infectious.  The proportional change of node $i$ is calculated as the number of agents that are infected when node $i$ has been removed divided by the number of nodes that were originally infected, and then that subtracted from one so that a value of one means that no nodes become infected if this one is removed.  Alternatively the delta fraction is the fraction of infected nodes in the original run minus the fraction of nodes that become infected when node $i$ is removed.  For both versions negative values occur when more nodes become infected contingent upon $i$'s removal compared to the original run. An agent that was never infected will have a TKO value of zero for all its temporal nodes.  For each of these temporal node-based measures we aggregate them to agents by considering both the maximum value an agent achieves across time and its average TKO score across time.

The Pearson correlations for agent TKO scores and magnitude appear in Table \ref{MagnitudeVsTKOPearson}; in the most correlated scenario (SIR smallworld 0.10 infection rate) the best match is to maximum TKO with a correlation coefficient just under 0.50 (marked with *).  Although we initially believed that the Spearman rank correlations would be higher, they are actually very similar and not consistently better or worse (a table of Spearman correlations appears in the Supplementary Materials).  For example, the best-case scenario for the Spearman correlation is the same, with a Spearman rho value of 0.51658.  For both types of correlation the performance drops dramatically as the disease magnitude increases (via higher infection rates), indicating that the large proportion of runs with almost no spread (``duds'') are trivially improving the correlations and overstating the ability of agent-initiated magnitude to measure propagation impact.

\begin{center} \begin{table}[!ht] \centering
\scalebox{ 0.78 }{
\rowcolors{1}{white}{tableShade}
\begin{tabular}{ccc d{1.6} d{1.6} d{1.6} d{1.6}} \toprule
 \text{Disease Type} & \text{Network Type} & \multicolumn{1}{c}{InfectionRate} & \multicolumn{1}{c}{MaxProportion} & \multicolumn{1}{c}{MaxDeltaFraction} & \multicolumn{1}{c}{AveProportion} &
  \multicolumn{1}{c}{AveDeltaFraction}  \\ \hline
 \text{SIR} & \text{scalefree} & 0.10 & 0.402831 & 0.404905 & 0.288126 & 0.292081 \\
 \text{SIR} & \text{scalefree} & 0.15 & 0.0674457 & 0.246309 & 0.0644679 & 0.157114  \\
 \text{SIR} & \text{scalefree} & 0.20 & 0.0457449 & 0.219299 & 0.0781154 & 0.158489 \\ \hline
 \text{SIR} & \text{smallworld} & 0.10 & 0.494424* & 0.471569 & 0.366466 & 0.363984  \\
 \text{SIR} & \text{smallworld} & 0.15 & 0.043098 & 0.264589 & 0.0770296 & 0.188738 \\
 \text{SIR} & \text{smallworld} & 0.20 & 0.0297556 & 0.19209 & 0.0154106 & 0.118696 \\ \hline
 \text{SIS} & \text{scalefree} & 0.10 & 0.346553 & 0.375933 & 0.268197 & 0.282026  \\
 \text{SIS} & \text{scalefree} & 0.15 & 0.0566606 & 0.247974 & 0.0838702 & 0.153358 \\
 \text{SIS} & \text{scalefree} & 0.20 & 0.0446214 & 0.233998 & 0.0585834 & 0.107849  \\ \hline
 \text{SIS} & \text{smallworld} & 0.10 & 0.404204 & 0.417726 & 0.353329 & 0.370916  \\
 \text{SIS} & \text{smallworld} & 0.15 & 0.0240389 & 0.201376 & 0.0452133 & 0.149954  \\
 \text{SIS} & \text{smallworld} & 0.20 & 0.0502245 & 0.190489 & 0.0420603 & 0.10682  \\
\specialrule{.08em}{.00em}{.00em} 
\end{tabular} } \caption{The mean Pearson correlations coefficients across the 25 network instantiations of the disease magnitude given an agent is the initially infected agents and the TKO scores for that agent.  The low correlations imply that using the disease spread based on initial infection is a poor measure of influence.} \label{MagnitudeVsTKOPearson} 
\end{table} \end{center} \vspace{-10mm}

\begin{figure*}[!ht]
\centering
\includegraphics[width=0.9\textwidth]{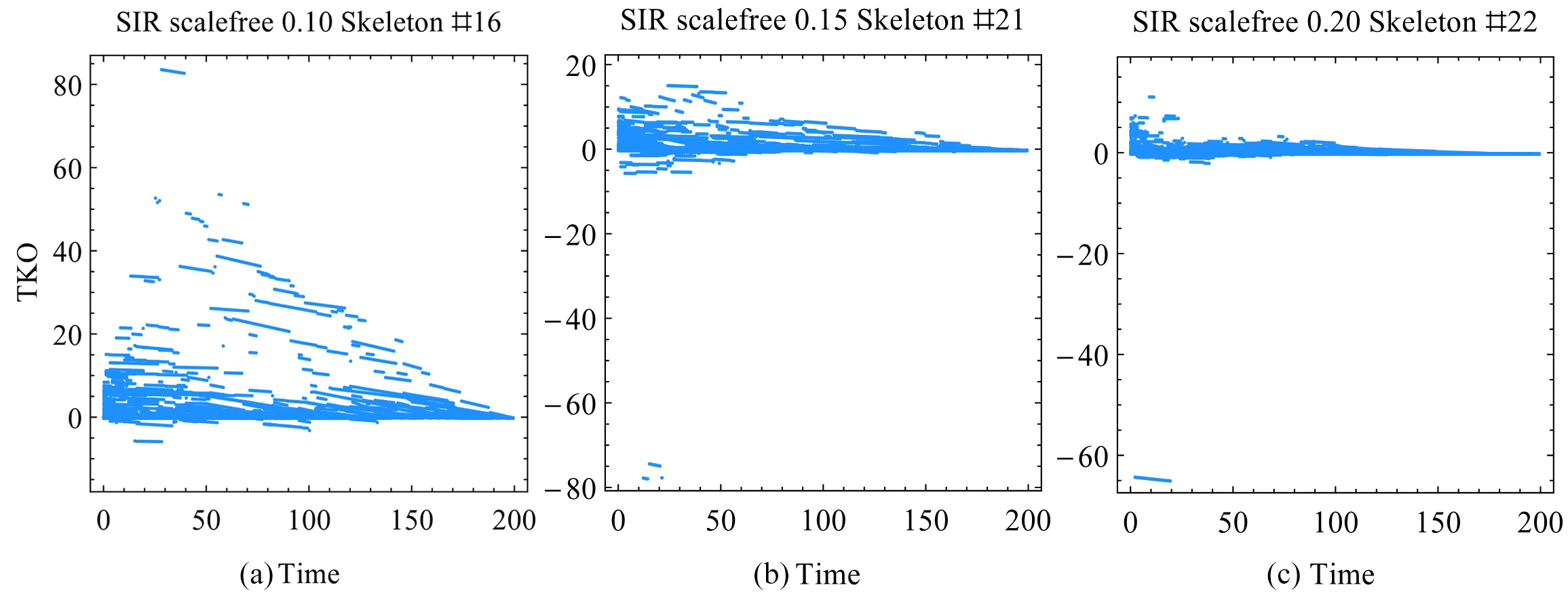}
\caption{Plot of TKO scores across time for SIR dynamics and a scalefree network.  These examples shows that the most influential agent-times often do not occur during the initial phases of a disease, but can indicate bottlenecks in the spread of the disease. This also shows the appearance of negative TKO agents, the removal of which actually \i{increases} the morbidity of the disease.}\label{TkoAcrossTime}
\end{figure*}

The poor correlations between TKO and agent-initiated magnitude have multiple explanations.  To understand the relationship better we present a few select plots of the agent TKO scores across time in Figure \ref{TkoAcrossTime}.  These plots present the change in magnitude resulting from removing each infectious agent at each time averaged across the 200 runs initialized with each agent being infected.  So a value of $m$ means that \i{on average} removing this agent at this time decreases morbidity by $m$ agent-times.  As we saw, there are many dud runs in which the disease doesn't spread beyond a few initial agents; such cases bring down the average values.  A TKO score of twenty might mean 500 saved agent-times in one run and none in the others, or 50 in ten runs, etc.  So TKO scores can be small if the disease tends not to spread much because no agent at no time will be a key player in the localized infections.  On the other hand, when the infection rate is 0.20 the disease spreads to many more agents across time, enough that no single agent could be responsible for the scale of the infection across multiple initializations.  There are just too many infection paths for any one agent to be a key player on enough of them to have a high knockout effect.

Up to this point we have argued that temporal magnitude is more accurate than cumulative cases as a measure of disease morbidity because it accounts for variations in the length of infection and also reinfection.  A network measure's ability to capture an agent's influence on disease is standardly compared to the eventual spread of the disease contingent upon it starting at that agent, but our analysis of correlations with TKO shows that this standard measure of impact itself fails to capture how much disease spread that agent is responsible for because it lacks sensitivity to the structure of the interactions across time.  From these results we tentatively conclude that TKO stands as the best measure of an agent's influence on network propagation.  We now turn to testing the ability of static network measures to identify a system's high-impact agents.

\subsection{Predicting Temporal Knockout from the Static Interaction Network}

The temporally extruded network structure captures the system dynamics in a way that facilitates contingency analyses, however one must already have the data across time to measure those properties, including TKO.  For predictive purposes we would like to know if there is some property of the known interaction structure that can identify key players \citep{yu2010finding}.  Although temporal networks are gaining popularity (see \citep{holme2015modern} for a review), most network analysis is still performed on flat networks because there are already measures available with known interpretations.  The question here is whether any flat graph property can accurately predict the conditional marginal infection as measured by agent-aggregated temporal knockout.

We ran the three comparisons between the four aggregated TKO measures five network centrality measures.  Both Pearson and Spearman correlations were calculated.   Furthermore, because the standard network centrality measures only purport to capture the highest value agents properly; i.e., rather than a claim to assigning accurate values to all nodes we also compared the overlap between the ten agents (5\%) with the top TKO scores with the ten agents with the top centrality scores \cite{Kitsak2010}.  We compared the maximum proportional and maximum delta fraction TKO as well as the average proportional and average delta fraction TKOs with degree, closeness, betweenness, eigenvector, and Katz centrality ($k$-core values were too undifferentiated on our base networks to be meaningful).  The full output of the analysis appears in the Supplementary Materials.

\begin{center} \begin{table}[!ht] \centering
\scalebox{ 0.83 }{
\rowcolors{2}{}{tableShade}
\begin{tabular}{ccc d{1.6} d{1.6} d{1.6} d{1.6} d{1.6}}
 \specialrule{.08em}{.00em}{.00em} 
 \text{Disease Type} & \text{Network Type} &\multicolumn{1}{c}{InfectionRate} & \multicolumn{1}{c}{Degree} & \multicolumn{1}{c}{Closeness} & \multicolumn{1}{c}{Betweenness} & \multicolumn{1}{c}{Eigenvector} & \multicolumn{1}{c}{Katz}  \\ \hline

 \text{SIR} & \text{scalefree} & 0.10 & 0.12721 & 0.10583 & 0.09908 & 0.09757 & 0.07997 \\
 \text{SIR} & \text{scalefree} & 0.15 & 0.12514 & 0.09557 & 0.10187 & 0.0977 & 0.08398 \\
 \text{SIR} & \text{scalefree} & 0.20 & 0.13992 & 0.11872 & 0.11503 & 0.11485 & 0.10137 \\ \hline
 \text{SIR} & \text{smallworld} & 0.10 & -0.00492 & 0.03076 & 0.01898 & -0.00766 & -0.01483 \\
 \text{SIR} & \text{smallworld} & 0.15 & 0.04866 & 0.07044 & 0.08208 & 0.08047 & 0.06403 \\
 \text{SIR} & \text{smallworld} & 0.20 & 0.06674 & 0.12695 & 0.10153 & 0.02619 & 0.05665 \\ \hline
 \text{SIS} & \text{scalefree} & 0.10 & 0.13528 & 0.08851 & 0.11109 & 0.09088 & 0.06804 \\
 \text{SIS} & \text{scalefree} & 0.15 & 0.16164 & 0.09822 & 0.127 & 0.10646 & 0.08486 \\
 \text{SIS} & \text{scalefree} & 0.20 & 0.23309 & 0.16747 & 0.19103 & 0.16756 & 0.13834 \\ \hline
 \text{SIS} & \text{smallworld} & 0.10 & 0.02536 & 0.0545 & 0.01982 & -0.0126 & 0.0042 \\
 \text{SIS} & \text{smallworld} & 0.15 & 0.06834 & 0.10208 & 0.11029 & 0.02545 & 0.05613 \\
 \text{SIS} & \text{smallworld} & 0.20 & 0.12315 & 0.17406 & 0.24128* & 0.0522 & 0.12552 \\

\specialrule{.08em}{.00em}{.00em} 
\end{tabular} } \caption{The Pearson correlations between the mean proportional TKO score with each of five base network agent centrality scores.  Other results tables appear in the Supplementary Materials.} 
\end{table} \end{center} \vspace{-10mm}

We find that neither the Pearson or the Spearman correlations are systematically higher, nor is any one of the network measures consistently better than all the others (although eigenvector and Katz centrality are consistently worse).  Although the correlations are typically positive, the correlation coefficients and Spearman Rhos are almost entirely below 0.20.  Differences between the proportional and delta fraction TKOs are small (as expected), but not negligible; delta fractional correlations tend to be better but not in every case.  Similarly, the correlations with mean TKO tend to be slightly higher than maximum TKO, but the differences are small and inconsistent.  For the top ten overlap comparison we find that the centrality measures only rarely manage to find one of the top ten TKO agents; eigenvector and Katz centrality never do.

There are other patterns in the results that may offer clues to where to look for improved network measures.  For example, for each disease type, each network type, and each TKO version the correlations of all measures tend to be higher with larger infection rates.  Unsurprisingly, degree centrality typically performs better on the scalefree networks than the small world networks.  However, any such pattern may be spurious because the correlation values are too low and similar for our sample size to provide adequate power.  In summary, the result is that none of the five centrality measures on the flat interaction network can predict which agents have the greatest influence on spreading a disease.

\section{Conclusions}

In this paper we have argued that using temporal networks to capture disease spread has the benefits of  incorporating the details of the interaction timing which is necessary for judging each agent's level of influence/impact on the spread.  The number of infectious agent-time nodes, a measure we call magnitude, is a superior to cumulative cases because it captures both the length of infections and reinfection.    However, adapting the standard measures of influence -- eventual spread contingent upon the starting agent or blocked spread contingent upon removing the agent -- to magnitude is insufficient to properly capture an agent's overall level of influence.  Although eliminating the initial agent is a sure-fire way to stop the spread, that is not informative for deciding whom to remove before the disease starts.  What is needed is the change in the spread of disease contingent upon each agent being removed generalized over all possible initial agents.  But the degree of influence is also dependent on \i{when} the agent is removed because the interaction dynamics of these systems are complex: removing an agent early can increase the eventual spread.  We present the temporal knockout measure to capture all these contingencies and provide a general benchmark for propagation influence.

One key insight from this study is that an agent's influence depends on how the dynamics unfold through time, which cannot be accurately predicted by historic interaction data or known communication channels.  Nascent measures on temporal network structure (i.e., ones that operate on the full temporal web) can accurately track the TKO property with considerably less computational time, but they still require knowing the complete interaction structure over time \cite{bramson2015temporal}.  Thus, they work as effective \i{proxy} measures, but are not viable \i{predictor} measures.  Although we do not have improved static network measures to offer at this stage, we believe that having a proper benchmark for such measures provides the foundation necessary for developing them.

For most realistic health applications, by the time an intervention occurs there are already several infectious individuals, and for this reason there is interest in measures/strategies for scenarios with multiple initially infected agents \citep{danon2011networks}.  The problem is in the combinatorics; e.g., instead of 200 runs per network, with two initial agents it becomes ${200 \choose 2} = 19,900$ runs -- for just three initial agents it becomes 1,313,400 runs.  Because TKO generalizes marginal conditional spread of every agent-time across all initially infected agents, the TKOs scores can be combined post hoc without needing to rerun the simulations. So, although the TKO algorithm is computationally intense compared to the single initial agent runs, There would be considerable time savings when compared to testing every combination of initially infected agents.  

As noted by \cite{Kitsak2010}, when using cumulative cases to capture the influence of particular agents it makes sense to keep the infection probabilities small enough that the disease typically will not spread to the whole population -- otherwise the role of any single individual will be difficult to discern.  TKO does not suffer from this limitation because the disease magnitude measure also detects delays in infection even if the whole population does eventually get infected.  Again, the timing of the interactions is important, so in addition to facilitating a reduction in morbidity, TKO is useful for developing adaptive intervention strategies.

Recent papers have introduced new measures with claims of increased accuracy (at least in certain contexts).  However, those accuracy claims are based on how well their own measure matched their own chosen metric on their own chosen network and spread parameters.  We propose that TKO, in its exhaustive marginal contingent effect calculation, can act as a benchmark metric against which the performance of proposed measures can be judged -- essentially establishing a ground truth for the influence of each agent (at each time) in a network.  



We acknowledge that the version of temporal knockout presented here is not the only option for benchmarking epidemiological network studies.  One direction to look for further improvement is increasing the refinement of the measures through, for example, another layer of contingency.  Another direction is to expand the breadth of the simulations to more closely approach an exhaustive analysis of interaction possibilities.  We visit these ideas in follow-up research to establish shared benchmarks for evaluating measures of network influence on a variety of standardized networks similar to how Zachary's Karate Club has been used to test community detection methods.  Before such benchmark networks can be established, we as a community must agree on what counts as a measure of influence.  We propose that temporal knock out may fill that role, and at the very least is a useful step in the right direction.


\section*{Acknowledgements}
We would like to thank Koen Schoors for making this collaboration possible.
This work is supported by the Research Foundation Flanders (FWO-Flanders).

\bibliography{TemporalWebRefs}

\clearpage
\section*{Supplementary Material}

\noindent \b{Model Scenarios and Infection Sizes: } Includes a 3D histogram with a row for each scenario showing the frequency of infections of each size.  We also have a set of twelve 3D histograms (one for each scenario) with a row for each of the 25 skeletons showing the disease variation resulting from network structure; however, in consideration of space and the real focus of this paper they are excluded (available upon request).  We also provide a table of the mean and standard deviations of the raw magnitudes for each scenario.  When excluding the duds the distributions approximate normal distributions, but the large numbers of duds make the normal approximation inappropriate and it is \i{not} the case that the disease results follow any single distribution with the mean and standard deviations in the table.  The mean and standard deviation do, however, capture the relative all-things-considered expected infection sizes for each scenario. \\

\noindent \b{Correlations between Agent-Initialized Magnitude and TKO Measures: } The Pearson and Spearman correlation coefficients between (a) the disease magnitude reached when agent $i$ is the initial agent and (b) four different versions of the TKO aggregated across time for agent $i$.  The correlations are performed separately for each scenario between the lists of values for all 200 agents combined across all 25 network skeletons.  With the highest scores near 0.50 and most much lower, the result is that measuring an agent's impact using the super-spreader approach alone is not accurate in capturing an agent's actual influence compared to TKO.\\

\noindent \b{Comparisons of Network Measures to TKO scores: } This section starts with one page of further methodological description, especially about the flattened observed interaction dynamics networks.  Following that are eight pages of table triplets each showing the Pearson, Spearman, and Top Ten comparisons between each of five common network centrality measures.  Each of the four TKO variation has it's one page of tables for both the base and the flattened networks.  Because the base and unweighted flattened networks are nearly identical, so are the correlations.  The weighted versions of the flattened measure are excluded in consideration of the space to describe them in light of the result that they also do not significantly covary with any TKO measures.

\clearpage

\subsection*{Model Scenarios and Infection Sizes}
 \vspace{-5mm}
 \vspace{-5mm}

\begin{center} \begin{table}[!ht] \centering
\scalebox{ 0.78 }{
\rowcolors{1}{white}{tableShade}
\begin{tabular}{c c c d{5.5} d{5.5} d{5.5}}\toprule
Infection Type & Network Type & Infection Probability  &  \multicolumn{1}{c}{Mean Magnitude}  &   \multicolumn{1}{c}{Magnitude StDev} &  \multicolumn{1}{c}{Percent Duds}\\  \hline
SIR & Scale Free &  0.10 & 143.352 & 288.549 & 0.625 \\
SIR & Scale Free &  0.15 & 584.744 & 774.628 & 0.482 \\
SIR & Scale Free &  0.20 & 1296.44 & 1142.24 & 0.380 \\ \hline
SIR & Small World &  0.10 & 88.9266 & 131.743 & 0.584 \\
SIR & Small World &  0.15 & 227.321 & 324.207 & 0.457 \\
SIR & Small World &  0.20 & 445.033 & 559.017 & 0.352 \\ \hline
SIS & Scale Free &  0.10 & 548.746 & 1155.19 & 0.593 \\
SIS & Scale Free &  0.15 & 5003.03 & 5237.44 & 0.445 \\
SIS & Scale Free &  0.20 & 10800.6 & 8150.76 & 0.344 \\ \hline
SIS & Small World &  0.10 & 308.734 & 536.106 & 0.557 \\
SIS & Small World &  0.15 & 2526.97 & 2839.26 & 0.433 \\
SIS & Small World &  0.20 & 7036.79 & 5623.18 & 0.333 \\
\specialrule{.08em}{.00em}{.00em} 
\end{tabular} }
\caption{Results summary of infection spread for each model variation.  Each row aggregates 5000 runs (one run initialized at each of 200 agents for each of the 25 base network implementations). Duds are defined as runs in which the raw magnitude is fewer than 50 agent-times. } \label{MagnitudeSummary} 
\end{table} \end{center} \vspace{-10mm}

\clearpage
\subsection*{Correlations between Agent-Initialized Magnitude and TKO Measures}

\begin{center} \begin{table}[!ht] \centering
\scalebox{ 0.78 }{
\rowcolors{2}{tableShade}{}
\begin{tabular}{ccc d{1.6} d{1.6} d{1.6} d{1.6}} 
\multicolumn{7}{c}{ \text{Pearson Correlations of Agent-Initialized Magnitude and TKO Measures.}} \\
\specialrule{.08em}{.00em}{.00em} 
 \text{Disease Type} & \text{Network Type} & \multicolumn{1}{c}{InfectionRate} & \multicolumn{1}{c}{MaxProportion} & \multicolumn{1}{c}{MaxDeltaFraction} & \multicolumn{1}{c}{AveProportion} &
  \multicolumn{1}{c}{AveDeltaFraction}  \\ \hline
 \text{SIR} & \text{scalefree} & 0.10 & 0.402831 & 0.404905 & 0.288126 & 0.292081 \\
 \text{SIR} & \text{scalefree} & 0.15 & 0.0674457 & 0.246309 & 0.0644679 & 0.157114  \\
 \text{SIR} & \text{scalefree} & 0.20 & 0.0457449 & 0.219299 & 0.0781154 & 0.158489 \\ \hline
 \text{SIR} & \text{smallworld} & 0.10 & 0.494424* & 0.471569 & 0.366466 & 0.363984  \\
 \text{SIR} & \text{smallworld} & 0.15 & 0.043098 & 0.264589 & 0.0770296 & 0.188738 \\
 \text{SIR} & \text{smallworld} & 0.20 & 0.0297556 & 0.19209 & 0.0154106 & 0.118696 \\ \hline
 \text{SIS} & \text{scalefree} & 0.10 & 0.346553 & 0.375933 & 0.268197 & 0.282026  \\
 \text{SIS} & \text{scalefree} & 0.15 & 0.0566606 & 0.247974 & 0.0838702 & 0.153358 \\
 \text{SIS} & \text{scalefree} & 0.20 & 0.0446214 & 0.233998 & 0.0585834 & 0.107849  \\ \hline
 \text{SIS} & \text{smallworld} & 0.10 & 0.404204 & 0.417726 & 0.353329 & 0.370916  \\
 \text{SIS} & \text{smallworld} & 0.15 & 0.0240389 & 0.201376 & 0.0452133 & 0.149954  \\
 \text{SIS} & \text{smallworld} & 0.20 & 0.0502245 & 0.190489 & 0.0420603 & 0.10682  \\
\specialrule{.08em}{.00em}{.00em} 
\end{tabular} } \caption{The mean Pearson correlations coefficients of (a) the disease magnitude given an agent is the initially infected agent and (b) the TKO score for that agent.  The on-average low correlations imply that using the disease spread based on initial infection is a poor measure of influence.  Furthermore, the correlations are nearly always worse with increasing infection rates (and hence increasing magnitudes and fewer dud runs) implying that much of the ability to match TKO relies on the cases in which both scores are near zero. } 
\end{table} \end{center} \vspace{-10mm}

\begin{center} \begin{table}[!ht] \centering
\scalebox{ 0.78 }{
\rowcolors{2}{tableShade}{}
\begin{tabular}{ccc d{1.6} d{1.6} d{1.6} d{1.6}}
\multicolumn{7}{c}{ \text{Spearman Correlations of Agent-Initialized Magnitude and TKO Measures.}} \\
\specialrule{.08em}{.00em}{.00em} 
 \text{Disease Type} & \text{Network Type} &\multicolumn{1}{c}{InfectionRate} & \multicolumn{1}{c}{MaxProportion} & \multicolumn{1}{c}{MaxDeltaFraction} & \multicolumn{1}{c}{AveProportion} & \multicolumn{1}{c}{AveDeltaFraction}  \\ \hline
 \text{SIR} & \text{scalefree} & 0.10 &  0.40153 & 0.305474 & 0.388263 & 0.308888 \\
 \text{SIR} & \text{scalefree} & 0.15 &  0.0630928 & 0.221007 & 0.105923 & 0.193733 \\
 \text{SIR} & \text{scalefree} & 0.20 &  0.0277719 & 0.203116 & 0.0667476 & 0.168757 \\ \hline
 \text{SIR} & \text{smallworld} & 0.10 & 0.516584* & 0.420064 & 0.461123 & 0.404163 \\
 \text{SIR} & \text{smallworld} & 0.15 & 0.0671833 & 0.292652 & 0.149355 & 0.270376 \\
 \text{SIR} & \text{smallworld} & 0.20 & 0.0375412 & 0.222854 & 0.0577158 & 0.171541 \\ \hline
 \text{SIS} & \text{scalefree} & 0.10 & 0.296555 & 0.252962 & 0.277245 & 0.235637 \\
 \text{SIS} & \text{scalefree} & 0.15 & 0.0487628 & 0.198384 & 0.0806729 & 0.13593 \\
 \text{SIS} & \text{scalefree} & 0.20 & 0.0370982 & 0.184936 & 0.027529 & 0.069931 \\ \hline
 \text{SIS} & \text{smallworld} & 0.10 & 0.335403 & 0.280811 & 0.29627 & 0.255873 \\
 \text{SIS} & \text{smallworld} & 0.15 & 0.0256655 & 0.190925 & 0.0797812 & 0.160578 \\
 \text{SIS} & \text{smallworld} & 0.20 & 0.0449195 & 0.187943 & 0.0447358 & 0.11443 \\
\specialrule{.08em}{.00em}{.00em} 
\end{tabular} } \caption{The mean Spearman Rank correlation coefficients (rho) of (a) the disease magnitude given an agent is the initially infected agent and (b) the TKO score for that agent.  The correlations reveal similar values and a similar pattern to the Pearson correlations, reinforcing that using the disease spread based on initial infection is a poor measure of influence.}
\end{table} \end{center} \vspace{-10mm}

\clearpage

\subsection*{Comparisons of Network Measures to TKO scores}

The following twelve sets of three data tables present the results of determining how well common network centrality measures capture agent influence.  Although the main result is that none of the network measures successfully capture/predict agent influence as measured by four versions of TKO in any scenario, the specific changes in the data reveal patterns -- and those patterns may point to improved measures.  

Although the paper focuses on the base network analysis, we also analyzed the network generated by flattening the observed interactions.  We record who interacts with whom over time in the temporal network skeleton, then we flatten this skeleton to achieve both a weighted by interaction frequency and an unweighted flat network representation.   If the model runs long enough the observed interactions converge to the base network of potential interactions, but in many applications the flattened network is observable/derivable from data while the base network is unknown and/or theoretical.  In our simulations, because the probability that a given link is active in a time step is $\propto 1/k$, $k$ is low (typically single digit except a few agents in the scale free networks), and there are 200 time steps, the base and unweighted flattened graphs are nearly identical.  

Because for each base network we generate the skeleton including all transition and interaction probabilities, the empirically derived flattened network connections are always the same for each run of the same skeleton (i.e., starting from each agent).  In the current model the infection state does not alter the interaction probability.  If it did, then the observed transitions would vary from run to run even using the same network skeleton because what is stored in the skeleton is a set of draws from probability distributions rather than a fixed interaction structure.  If, for example, being infectious reduced the probability of interaction, then the probability stored in the skeleton would be compared to a different interaction threshold and thus could alter which interactions occur.  However, using the same skeletons for multiple runs of different dynamics on the same structure at least satisfies the Markov condition for these simulations, which is not maintained when running the dynamics independently for each initial agent run.

Flattened graphs are potentially better at tracking influence because they allow one to create weighted networks from the observed interaction frequencies.  However, in our experiments the correlation valued between TKO and the weighted network centrality measures were no better, although they were slightly different.  For this reason and considerations of space we have excluded them from this paper.

\clearpage
\begin{center}\large{Maximum TKO and Base Interaction Network}\end{center}

\begin{center} \begin{table}[!ht] \centering
\scalebox{ 0.83 }{
\rowcolors{2}{tableShade}{}
\begin{tabular}{ccc d{1.6} d{1.6} d{1.6} d{1.6} d{1.6}}
\multicolumn{8}{c}{ \text{Pearson Correlations of Centrality Measures and Maximum TKO on Base Network }} \\ \specialrule{.08em}{.00em}{.00em} 
 \text{Disease Type} & \text{Network Type} &\multicolumn{1}{c}{InfectionRate} & \multicolumn{1}{c}{Degree} & \multicolumn{1}{c}{Closeness} & \multicolumn{1}{c}{Betweenness} & \multicolumn{1}{c}{Eigenvector} & \multicolumn{1}{c}{Katz}  \\ \hline

 \text{SIR} & \text{scalefree} & 0.10 & 0.12254 & 0.09622 & 0.09557 & 0.08855 & 0.06909 \\
 \text{SIR} & \text{scalefree} & 0.15 & 0.1475 & 0.11315 & 0.127 & 0.11531 & 0.09876 \\
 \text{SIR} & \text{scalefree} & 0.20 & 0.14354 & 0.11432 & 0.11925 & 0.11355 & 0.09922 \\  \hline
 \text{SIR} & \text{smallworld} & 0.10 & 0.00727 & 0.0654 & 0.03754 & 0.00047 & -0.0051 \\
 \text{SIR} & \text{smallworld} & 0.15 & 0.05936 & 0.0854 & 0.11007 & 0.08465 & 0.0749 \\
 \text{SIR} & \text{smallworld} & 0.20 & 0.08282 & 0.17901 & 0.1399 & 0.03241 & 0.07405 \\ \hline
 \text{SIS} & \text{scalefree} & 0.10 & 0.13835 & 0.09571 & 0.11868 & 0.09935 & 0.07816 \\
 \text{SIS} & \text{scalefree} & 0.15 & 0.11807 & 0.08243 & 0.09698 & 0.0866 & 0.07341 \\
 \text{SIS} & \text{scalefree} & 0.20 & 0.09523 & 0.07743 & 0.07644 & 0.07262 & 0.06029 \\  \hline
 \text{SIS} & \text{smallworld} & 0.10 & 0.02154 & 0.07993 & 0.03546 & -0.01334 & 0.00489 \\
 \text{SIS} & \text{smallworld} & 0.15 & 0.0637 & 0.11582 & 0.11384 & 0.02679 & 0.05537 \\
 \text{SIS} & \text{smallworld} & 0.20 & 0.11755 & 0.13727 & 0.22958 & 0.06551 & 0.13365 \\  

\specialrule{.08em}{.00em}{.00em} 
\rowcolor{white}
\multicolumn{8}{c}{ \text{  }}\\
\multicolumn{8}{c}{ \text{  }}\\ 
\rowcolor{white}
\multicolumn{8}{c}{ \text{Spearman Correlations of Centrality Measures and Maximum TKO on Base Network }} \\ \specialrule{.08em}{.00em}{.00em} 
 \text{Disease Type} & \text{Network Type} &\multicolumn{1}{c}{InfectionRate} & \multicolumn{1}{c}{Degree} & \multicolumn{1}{c}{Closeness} & \multicolumn{1}{c}{Betweenness} & \multicolumn{1}{c}{Eigenvector} & \multicolumn{1}{c}{Katz}  \\ \hline

 \text{SIR} & \text{scalefree} & 0.10 & 0.19228 & 0.10423 & 0.18884 & 0.09346 & 0.06062 \\
 \text{SIR} & \text{scalefree} & 0.15 & 0.19853 & 0.10738 & 0.17779 & 0.10471 & 0.07401 \\
 \text{SIR} & \text{scalefree} & 0.20 & 0.21217 & 0.12175 & 0.20802 & 0.11573 & 0.08827 \\  \hline
 \text{SIR} & \text{smallworld} & 0.10 & 0.01237 & 0.02756 & 0.03768 & 0.03103 & 0.01127 \\
 \text{SIR} & \text{smallworld} & 0.15 & 0.05613 & 0.09342 & 0.09055 & 0.06225 & 0.04637 \\
 \text{SIR} & \text{smallworld} & 0.20 & 0.08772 & 0.17172 & 0.10234 & 0.07409 & 0.08572 \\  \hline
 \text{SIS} & \text{scalefree} & 0.10 & 0.18963 & 0.0922 & 0.18953 & 0.08388 & 0.0518 \\
 \text{SIS} & \text{scalefree} & 0.15 & 0.18795 & 0.08439 & 0.17362 & 0.08274 & 0.05498 \\
 \text{SIS} & \text{scalefree} & 0.20 & 0.19385 & 0.08551 & 0.19045 & 0.08034 & 0.04848 \\  \hline
 \text{SIS} & \text{smallworld} & 0.10 & 0.01626 & 0.04382 & 0.02805 & -0.00418 & -0.00617 \\
 \text{SIS} & \text{smallworld} & 0.15 & 0.08308 & 0.15921 & 0.14567 & 0.03365 & 0.05083 \\
 \text{SIS} & \text{smallworld} & 0.20 & 0.11411 & 0.16911 & 0.20342 & 0.09623 & 0.13965 \\ 
 
\specialrule{.08em}{.00em}{.00em} 
\rowcolor{white}
\multicolumn{8}{c}{ \text{  }}\\
\multicolumn{8}{c}{ \text{  }}\\
\rowcolor{white}
\multicolumn{8}{c}{ \text{Top Ten Overlap of Centrality Measures and Maximum TKO on Base Network }} \\ \specialrule{.08em}{.00em}{.00em} 
 \text{Disease Type} & \text{Network Type} &\multicolumn{1}{c}{InfectionRate} & \multicolumn{1}{c}{Degree} & \multicolumn{1}{c}{Closeness} & \multicolumn{1}{c}{Betweenness} & \multicolumn{1}{c}{Eigenvector} & \multicolumn{1}{c}{Katz}  \\ \hline

 \text{SIR} & \text{scalefree} & 0.10 & 0.004 & 0.012 & 0.008 & 0. & 0. \\
 \text{SIR} & \text{scalefree} & 0.15 & 0.004 & 0.008 & 0.004 & 0. & 0. \\
 \text{SIR} & \text{scalefree} & 0.20 & 0.008 & 0.004 & 0.016 & 0. & 0. \\  \hline
 \text{SIR} & \text{smallworld} & 0.10 & 0. & 0.008 & 0.004 & 0. & 0. \\
 \text{SIR} & \text{smallworld} & 0.15 & 0.004 & 0.004 & 0.02 & 0. & 0. \\
 \text{SIR} & \text{smallworld} & 0.20 & 0.004 & 0.004 & 0.008 & 0. & 0. \\  \hline
 \text{SIS} & \text{scalefree} & 0.10 & 0.008 & 0.012 & 0.012 & 0. & 0. \\
 \text{SIS} & \text{scalefree} & 0.15 & 0. & 0.004 & 0. & 0. & 0. \\
 \text{SIS} & \text{scalefree} & 0.20 & 0.008 & 0.012 & 0.004 & 0. & 0. \\  \hline
 \text{SIS} & \text{smallworld} & 0.10 & 0. & 0.008 & 0.012 & 0. & 0. \\
 \text{SIS} & \text{smallworld} & 0.15 & 0.016 & 0.004 & 0.012 & 0. & 0. \\
 \text{SIS} & \text{smallworld} & 0.20 & 0.02 & 0.008 & 0.028 & 0. & 0. \\

\specialrule{.08em}{.00em}{.00em} 
\end{tabular} } \caption{The Pearson and Spearman correlations as well as the average percent of matching Top Ten agents between the maximum proportional TKO score with each of five base network agent centrality scores.} 
\end{table} \end{center} \vspace{-10mm}

\clearpage
\begin{center}\large{Maximum TKO and Flattened Interaction Network - Unweighted}\end{center}

\begin{center} \begin{table}[!ht] \centering
\scalebox{ 0.83 }{
\rowcolors{2}{tableShade}{}
\begin{tabular}{ccc d{1.6} d{1.6} d{1.6} d{1.6} d{1.6}}
\multicolumn{8}{c}{ \text{Pearson Correlations of Unweighted Centrality Measures and Maximum TKO on Flattened Network }} \\ \specialrule{.08em}{.00em}{.00em} 
 \text{Disease Type} & \text{Network Type} &\multicolumn{1}{c}{InfectionRate} & \multicolumn{1}{c}{Degree} & \multicolumn{1}{c}{Closeness} & \multicolumn{1}{c}{Betweenness} & \multicolumn{1}{c}{Eigenvector} & \multicolumn{1}{c}{Katz}  \\ \hline

 \text{SIR} & \text{scalefree} & 0.10 & 0.12308 & 0.09402 & 0.09698 & 0.071 & -0.01539 \\
 \text{SIR} & \text{scalefree} & 0.15 & 0.14731 & 0.11036 & 0.1275 & 0.05439 & 0.01392 \\
 \text{SIR} & \text{scalefree} & 0.20 & 0.14399 & 0.11477 & 0.12008 & 0.03457 & 0.00323 \\\hline
 \text{SIR} & \text{smallworld} & 0.10 & 0.00727 & 0.0654 & 0.03754 & -0.03346 & -0.0077 \\
 \text{SIR} & \text{smallworld} & 0.15 & 0.05936 & 0.0854 & 0.11007 & 0.04161 & -0.00132 \\
 \text{SIR} & \text{smallworld} & 0.20 & 0.08282 & 0.17901 & 0.1399 & -0.02131 & 0.01624 \\\hline
 \text{SIS} & \text{scalefree} & 0.10 & 0.13867 & 0.09368 & 0.11945 & 0.07499 & 0.00385 \\
 \text{SIS} & \text{scalefree} & 0.15 & 0.11804 & 0.07891 & 0.09743 & 0.06998 & 0.02266 \\
 \text{SIS} & \text{scalefree} & 0.20 & 0.09568 & 0.07786 & 0.07724 & 0.02772 & 0.01391 \\\hline
 \text{SIS} & \text{smallworld} & 0.10 & 0.02154 & 0.07993 & 0.03546 & -0.00236 & -0.00627 \\
 \text{SIS} & \text{smallworld} & 0.15 & 0.0637 & 0.11582 & 0.11384 & 0.00063 & 0.01076 \\
 \text{SIS} & \text{smallworld} & 0.20 & 0.11755 & 0.13727 & 0.22958 & -0.03648 & 0.02426 \\
 
\specialrule{.08em}{.00em}{.00em} 
\rowcolor{white}
\multicolumn{8}{c}{ \text{  }}\\
\multicolumn{8}{c}{ \text{  }}\\
\rowcolor{white}
\multicolumn{8}{c}{ \text{Spearman Correlations of Unweighted Centrality Measures and Maximum TKO on Flattened Network }} \\ \specialrule{.08em}{.00em}{.00em} 
 \text{Disease Type} & \text{Network Type} &\multicolumn{1}{c}{InfectionRate} & \multicolumn{1}{c}{Degree} & \multicolumn{1}{c}{Closeness} & \multicolumn{1}{c}{Betweenness} & \multicolumn{1}{c}{Eigenvector} & \multicolumn{1}{c}{Katz}  \\ \hline

 \text{SIR} & \text{scalefree} & 0.10 & 0.19227 & 0.10395 & 0.18916 & 0.06804 & -0.01847 \\
 \text{SIR} & \text{scalefree} & 0.15 & 0.19851 & 0.10683 & 0.18072 & 0.06697 & 0.02382 \\
 \text{SIR} & \text{scalefree} & 0.20 & 0.2122 & 0.12236 & 0.20727 & 0.03882 & 0.00276 \\ \hline
 \text{SIR} & \text{smallworld} & 0.10 & 0.01237 & 0.02756 & 0.03768 & -0.05264 & -0.00279 \\
 \text{SIR} & \text{smallworld} & 0.15 & 0.05613 & 0.09342 & 0.09055 & 0.05751 & -0.00228 \\
 \text{SIR} & \text{smallworld} & 0.20 & 0.08772 & 0.17172 & 0.10234 & 0.02494 & 0.01547 \\ \hline
 \text{SIS} & \text{scalefree} & 0.10 & 0.18963 & 0.09154 & 0.19067 & 0.05854 & -0.00145 \\
 \text{SIS} & \text{scalefree} & 0.15 & 0.1879 & 0.08214 & 0.17613 & 0.06379 & 0.03914 \\
 \text{SIS} & \text{scalefree} & 0.20 & 0.19386 & 0.08589 & 0.18989 & 0.02634 & 0.01455 \\ \hline
 \text{SIS} & \text{smallworld} & 0.10 & 0.01626 & 0.04382 & 0.02805 & -0.01907 & -0.01421 \\
 \text{SIS} & \text{smallworld} & 0.15 & 0.08308 & 0.15921 & 0.14567 & 0.04287 & 0.00706 \\
 \text{SIS} & \text{smallworld} & 0.20 & 0.11411 & 0.16911 & 0.20342 & 0.0116 & -0.0002 \\
 
\specialrule{.08em}{.00em}{.00em} 
\rowcolor{white}
\multicolumn{8}{c}{ \text{  }}\\
\multicolumn{8}{c}{ \text{  }}\\
\rowcolor{white}
\multicolumn{8}{c}{ \text{Top Ten Overlap of Unweighted Centrality Measures and Maximum TKO on Flattened Network }} \\ \specialrule{.08em}{.00em}{.00em} 
 \text{Disease Type} & \text{Network Type} &\multicolumn{1}{c}{InfectionRate} & \multicolumn{1}{c}{Degree} & \multicolumn{1}{c}{Closeness} & \multicolumn{1}{c}{Betweenness} & \multicolumn{1}{c}{Eigenvector} & \multicolumn{1}{c}{Katz}  \\ \hline
 
 \text{SIR} & \text{scalefree} & 0.10 & 0.004 & 0.004 & 0.004 & 0. & 0. \\
 \text{SIR} & \text{scalefree} & 0.15 & 0. & 0.008 & 0.012 & 0. & 0. \\
 \text{SIR} & \text{scalefree} & 0.20 & 0.008 & 0. & 0.016 & 0. & 0. \\ \hline
 \text{SIR} & \text{smallworld} & 0.10 & 0. & 0.008 & 0.004 & 0. & 0. \\
 \text{SIR} & \text{smallworld} & 0.15 & 0.004 & 0.004 & 0.02 & 0. & 0. \\
 \text{SIR} & \text{smallworld} & 0.20 & 0.004 & 0.004 & 0.008 & 0. & 0. \\ \hline
 \text{SIS} & \text{scalefree} & 0.10 & 0.008 & 0.016 & 0.012 & 0. & 0. \\
 \text{SIS} & \text{scalefree} & 0.15 & 0.004 & 0.004 & 0. & 0. & 0. \\
 \text{SIS} & \text{scalefree} & 0.20 & 0.008 & 0.012 & 0.004 & 0. & 0. \\ \hline
 \text{SIS} & \text{smallworld} & 0.10 & 0. & 0.008 & 0.012 & 0. & 0. \\
 \text{SIS} & \text{smallworld} & 0.15 & 0.016 & 0.004 & 0.012 & 0. & 0. \\
 \text{SIS} & \text{smallworld} & 0.20 & 0.02 & 0.008 & 0.028 & 0. & 0. \\

\specialrule{.08em}{.00em}{.00em} 
\end{tabular} } \caption{The Pearson and Spearman correlations as well as the average percent of matching Top Ten agents between the maximum proportional TKO score with each of five flattened observed interaction network agent centrality scores.  } 
\end{table} \end{center} \vspace{-10mm}

\clearpage
\begin{center}\large{Maximum Delta Fraction TKO and Base Interaction Network}\end{center}

\begin{center} \begin{table}[!ht] \centering
\scalebox{ 0.83 }{
\rowcolors{2}{tableShade}{}
\begin{tabular}{ccc d{1.6} d{1.6} d{1.6} d{1.6} d{1.6}}
\multicolumn{8}{c}{ \text{Pearson Correlations of Centrality Measures and Maximum Delta TKO on Base Network }} \\ \specialrule{.08em}{.00em}{.00em} 
 \text{Disease Type} & \text{Network Type} &\multicolumn{1}{c}{InfectionRate} & \multicolumn{1}{c}{Degree} & \multicolumn{1}{c}{Closeness} & \multicolumn{1}{c}{Betweenness} & \multicolumn{1}{c}{Eigenvector} & \multicolumn{1}{c}{Katz}  \\ \hline

 \text{SIR} & \text{scalefree} & 0.10 & 0.10286 & 0.08344 & 0.08 & 0.07491 & 0.05781 \\
 \text{SIR} & \text{scalefree} & 0.15 & 0.16692 & 0.13195 & 0.15128 & 0.13926 & 0.12469 \\
 \text{SIR} & \text{scalefree} & 0.20 & 0.15331 & 0.12103 & 0.12757 & 0.1196 & 0.10376 \\ \hline
 \text{SIR} & \text{smallworld} & 0.10 & 0.00478 & 0.08652 & 0.04694 & -0.00663 & -0.00979 \\
 \text{SIR} & \text{smallworld} & 0.15 & 0.08566 & 0.1837 & 0.14683 & 0.07832 & 0.09028 \\
 \text{SIR} & \text{smallworld} & 0.20 & 0.11087 & 0.20727 & 0.16839 & 0.04081 & 0.10154 \\ \hline
 \text{SIS} & \text{scalefree} & 0.10 & 0.11816 & 0.08506 & 0.10179 & 0.08919 & 0.0728 \\
 \text{SIS} & \text{scalefree} & 0.15 & 0.13843 & 0.10067 & 0.11605 & 0.1063 & 0.09209 \\
 \text{SIS} & \text{scalefree} & 0.20 & 0.11433 & 0.08954 & 0.09371 & 0.08694 & 0.07255 \\ \hline
 \text{SIS} & \text{smallworld} & 0.10 & 0.01778 & 0.11559 & 0.04528 & -0.00982 & 0.00079 \\
 \text{SIS} & \text{smallworld} & 0.15 & 0.08102 & 0.23611 & 0.15822 & 0.04475 & 0.07326 \\
 \text{SIS} & \text{smallworld} & 0.20 & 0.14115 & 0.20109 & 0.26869 & 0.07395 & 0.15273 \\
 
\specialrule{.08em}{.00em}{.00em} 
\rowcolor{white}
\multicolumn{8}{c}{ \text{  }}\\
\multicolumn{8}{c}{ \text{  }}\\ 
\rowcolor{white}
\multicolumn{8}{c}{ \text{Spearman Correlations of Centrality Measures and Maximum Delta TKO on Base Network }} \\ \specialrule{.08em}{.00em}{.00em} 
 \text{Disease Type} & \text{Network Type} &\multicolumn{1}{c}{InfectionRate} & \multicolumn{1}{c}{Degree} & \multicolumn{1}{c}{Closeness} & \multicolumn{1}{c}{Betweenness} & \multicolumn{1}{c}{Eigenvector} & \multicolumn{1}{c}{Katz}  \\ \hline

 \text{SIR} & \text{scalefree} & 0.10 & 0.19207 & 0.10484 & 0.19034 & 0.09269 & 0.05831 \\
 \text{SIR} & \text{scalefree} & 0.15 & 0.22213 & 0.12246 & 0.21438 & 0.11981 & 0.08953 \\
 \text{SIR} & \text{scalefree} & 0.20 & 0.23221 & 0.13404 & 0.2258 & 0.1246 & 0.09275 \\ \hline
 \text{SIR} & \text{smallworld} & 0.10 & 0.01617 & 0.05484 & 0.0493 & 0.0346 & 0.00932 \\
 \text{SIR} & \text{smallworld} & 0.15 & 0.06147 & 0.17675 & 0.09661 & 0.07697 & 0.06284 \\
 \text{SIR} & \text{smallworld} & 0.20 & 0.09899 & 0.24107 & 0.14145 & 0.07399 & 0.08672 \\ \hline
 \text{SIS} & \text{scalefree} & 0.10 & 0.18397 & 0.08878 & 0.18672 & 0.07923 & 0.04705 \\
 \text{SIS} & \text{scalefree} & 0.15 & 0.22487 & 0.09828 & 0.21779 & 0.0964 & 0.06375 \\
 \text{SIS} & \text{scalefree} & 0.20 & 0.22593 & 0.09769 & 0.22199 & 0.08987 & 0.05166 \\ \hline
 \text{SIS} & \text{smallworld} & 0.10 & 0.01798 & 0.07497 & 0.04538 & -0.01973 & -0.0145 \\
 \text{SIS} & \text{smallworld} & 0.15 & 0.08985 & 0.26058 & 0.16032 & 0.05387 & 0.06764 \\
 \text{SIS} & \text{smallworld} & 0.20 & 0.13121 & 0.23591 & 0.23906 & 0.09482 & 0.14568 \\
 
\specialrule{.08em}{.00em}{.00em}  
\rowcolor{white}
\multicolumn{8}{c}{ \text{  }}\\
\multicolumn{8}{c}{ \text{  }}\\
\rowcolor{white}
\multicolumn{8}{c}{ \text{Top Ten Overlap of Centrality Measures and Maximum Delta TKO on Base Network }} \\ \specialrule{.08em}{.00em}{.00em} 
 \text{Disease Type} & \text{Network Type} &\multicolumn{1}{c}{InfectionRate} & \multicolumn{1}{c}{Degree} & \multicolumn{1}{c}{Closeness} & \multicolumn{1}{c}{Betweenness} & \multicolumn{1}{c}{Eigenvector} & \multicolumn{1}{c}{Katz}  \\ \hline

 \text{SIR} & \text{scalefree} & 0.10 & 0.008 & 0.012 & 0.008 & 0. & 0. \\
 \text{SIR} & \text{scalefree} & 0.15 & 0.016 & 0.004 & 0.016 & 0. & 0. \\
 \text{SIR} & \text{scalefree} & 0.20 & 0.004 & 0.004 & 0.012 & 0. & 0. \\ \hline
 \text{SIR} & \text{smallworld} & 0.10 & 0.004 & 0.008 & 0.004 & 0. & 0. \\
 \text{SIR} & \text{smallworld} & 0.15 & 0.02 & 0.008 & 0.024 & 0. & 0. \\
 \text{SIR} & \text{smallworld} & 0.20 & 0.004 & 0.008 & 0.02 & 0. & 0. \\ \hline
 \text{SIS} & \text{scalefree} & 0.10 & 0.016 & 0.004 & 0.016 & 0. & 0. \\
 \text{SIS} & \text{scalefree} & 0.15 & 0.008 & 0.004 & 0.004 & 0. & 0. \\
 \text{SIS} & \text{scalefree} & 0.20 & 0.008 & 0.016 & 0.008 & 0. & 0. \\ \hline
 \text{SIS} & \text{smallworld} & 0.10 & 0.008 & 0.016 & 0.016 & 0. & 0. \\
 \text{SIS} & \text{smallworld} & 0.15 & 0.008 & 0.004 & 0.024 & 0. & 0. \\
 \text{SIS} & \text{smallworld} & 0.20 & 0.028 & 0.012 & 0.024 & 0. & 0. \\

\specialrule{.08em}{.00em}{.00em} 
\end{tabular} } \caption{The Pearson and Spearman correlations as well as the average percent of matching Top Ten agents between the maximum change in fractional TKO score with each of five base network agent centrality scores.} 
\end{table} \end{center} \vspace{-10mm}

\clearpage
\begin{center}\large{Maximum Delta Fraction TKO and Flattened Interaction Network - Unweighted}\end{center}

\begin{center} \begin{table}[!ht] \centering
\scalebox{ 0.83 }{
\rowcolors{2}{tableShade}{}
\begin{tabular}{ccc d{1.6} d{1.6} d{1.6} d{1.6} d{1.6}}
\multicolumn{8}{c}{ \text{Pearson Correlations of Unweighted Centrality Measures and Maximum Delta TKO on Flattened Network }} \\ \specialrule{.08em}{.00em}{.00em} 
 \text{Disease Type} & \text{Network Type} &\multicolumn{1}{c}{InfectionRate} & \multicolumn{1}{c}{Degree} & \multicolumn{1}{c}{Closeness} & \multicolumn{1}{c}{Betweenness} & \multicolumn{1}{c}{Eigenvector} & \multicolumn{1}{c}{Katz}  \\ \hline

 \text{SIR} & \text{scalefree} & 0.10 & 0.10322 & 0.08126 & 0.08119 & 0.0504 & -0.00984 \\
 \text{SIR} & \text{scalefree} & 0.15 & 0.16637 & 0.12859 & 0.15112 & 0.06161 & 0.02626 \\
 \text{SIR} & \text{scalefree} & 0.20 & 0.15391 & 0.12173 & 0.12867 & 0.04181 & 0.00983 \\ \hline
 \text{SIR} & \text{smallworld} & 0.10 & 0.00478 & 0.08652 & 0.04694 & -0.03715 & -0.00275 \\
 \text{SIR} & \text{smallworld} & 0.15 & 0.08566 & 0.1837 & 0.14683 & 0.07864 & 0.00211 \\
 \text{SIR} & \text{smallworld} & 0.20 & 0.11087 & 0.20727 & 0.16839 & -0.0042 & -0.00451 \\ \hline
 \text{SIS} & \text{scalefree} & 0.10 & 0.1181 & 0.08238 & 0.102 & 0.06581 & 0.00915 \\
 \text{SIS} & \text{scalefree} & 0.15 & 0.13813 & 0.09611 & 0.11633 & 0.08351 & 0.02489 \\
 \text{SIS} & \text{scalefree} & 0.20 & 0.11505 & 0.09018 & 0.09494 & 0.03574 & 0.01906 \\ \hline
 \text{SIS} & \text{smallworld} & 0.10 & 0.01778 & 0.11559 & 0.04528 & -0.01955 & 0.00412 \\
 \text{SIS} & \text{smallworld} & 0.15 & 0.08102 & 0.23611 & 0.15822 & 0.03885 & 0.01473 \\
 \text{SIS} & \text{smallworld} & 0.20 & 0.14115 & 0.20109 & 0.26869 & -0.04363 & 0.01438 \\
 
\specialrule{.08em}{.00em}{.00em} 
\rowcolor{white}
\multicolumn{8}{c}{ \text{  }}\\
\multicolumn{8}{c}{ \text{  }}\\
\rowcolor{white}
\multicolumn{8}{c}{ \text{Spearman Correlations of Unweighted Centrality Measures and Maximum Delta TKO on Flattened Network }} \\ \specialrule{.08em}{.00em}{.00em} 
 \text{Disease Type} & \text{Network Type} &\multicolumn{1}{c}{InfectionRate} & \multicolumn{1}{c}{Degree} & \multicolumn{1}{c}{Closeness} & \multicolumn{1}{c}{Betweenness} & \multicolumn{1}{c}{Eigenvector} & \multicolumn{1}{c}{Katz}  \\ \hline

 \text{SIR} & \text{scalefree} & 0.10 & 0.19206 & 0.10443 & 0.19097 & 0.06511 & -0.02302 \\
 \text{SIR} & \text{scalefree} & 0.15 & 0.22211 & 0.12175 & 0.21708 & 0.07393 & 0.01792 \\
 \text{SIR} & \text{scalefree} & 0.20 & 0.23223 & 0.13514 & 0.22504 & 0.05252 & 0.00736 \\ \hline
 \text{SIR} & \text{smallworld} & 0.10 & 0.01617 & 0.05484 & 0.0493 & -0.05967 & 0.00591 \\
 \text{SIR} & \text{smallworld} & 0.15 & 0.06147 & 0.17675 & 0.09661 & 0.11312 & 0.00537 \\
 \text{SIR} & \text{smallworld} & 0.20 & 0.09899 & 0.24107 & 0.14145 & 0.02893 & 0.0099 \\ \hline
 \text{SIS} & \text{scalefree} & 0.10 & 0.18396 & 0.08785 & 0.18809 & 0.05192 & 0.00408 \\
 \text{SIS} & \text{scalefree} & 0.15 & 0.22485 & 0.09573 & 0.22064 & 0.0759 & 0.03789 \\
 \text{SIS} & \text{scalefree} & 0.20 & 0.22593 & 0.09808 & 0.22135 & 0.04268 & 0.02456 \\ \hline
 \text{SIS} & \text{smallworld} & 0.10 & 0.01798 & 0.07497 & 0.04538 & -0.01514 & -0.01234 \\
 \text{SIS} & \text{smallworld} & 0.15 & 0.08985 & 0.26058 & 0.16032 & 0.10784 & 0.0099 \\
 \text{SIS} & \text{smallworld} & 0.20 & 0.13121 & 0.23591 & 0.23906 & 0.00745 & -0.01646 \\
 
\specialrule{.08em}{.00em}{.00em} 
\rowcolor{white}
\multicolumn{8}{c}{ \text{  }}\\
\multicolumn{8}{c}{ \text{  }}\\
\rowcolor{white}
\multicolumn{8}{c}{ \text{Top Ten Overlap of Unweighted Centrality Measures and Maximum Delta TKO on Flattened Network }} \\ \specialrule{.08em}{.00em}{.00em} 
 \text{Disease Type} & \text{Network Type} &\multicolumn{1}{c}{InfectionRate} & \multicolumn{1}{c}{Degree} & \multicolumn{1}{c}{Closeness} & \multicolumn{1}{c}{Betweenness} & \multicolumn{1}{c}{Eigenvector} & \multicolumn{1}{c}{Katz}  \\ \hline
 
 \text{SIR} & \text{scalefree} & 0.10 & 0.008 & 0.012 & 0.012 & 0. & 0. \\
 \text{SIR} & \text{scalefree} & 0.15 & 0.012 & 0.004 & 0.016 & 0. & 0. \\
 \text{SIR} & \text{scalefree} & 0.20 & 0.004 & 0. & 0.012 & 0. & 0. \\ \hline
 \text{SIR} & \text{smallworld} & 0.10 & 0.004 & 0.008 & 0.004 & 0. & 0. \\
 \text{SIR} & \text{smallworld} & 0.15 & 0.02 & 0.008 & 0.024 & 0. & 0. \\
 \text{SIR} & \text{smallworld} & 0.20 & 0.004 & 0.008 & 0.02 & 0. & 0. \\ \hline
 \text{SIS} & \text{scalefree} & 0.10 & 0.016 & 0.008 & 0.016 & 0. & 0. \\
 \text{SIS} & \text{scalefree} & 0.15 & 0.008 & 0.004 & 0.004 & 0. & 0. \\
 \text{SIS} & \text{scalefree} & 0.20 & 0.012 & 0.008 & 0.012 & 0. & 0. \\ \hline
 \text{SIS} & \text{smallworld} & 0.10 & 0.008 & 0.016 & 0.016 & 0. & 0. \\
 \text{SIS} & \text{smallworld} & 0.15 & 0.008 & 0.004 & 0.024 & 0. & 0. \\
 \text{SIS} & \text{smallworld} & 0.20 & 0.028 & 0.012 & 0.024 & 0. & 0. \\

\specialrule{.08em}{.00em}{.00em} 
\end{tabular} } \caption{The Pearson and Spearman correlations as well as the average percent of matching Top Ten agents between the maximum change in fractional TKO score with each of five flattened observed interaction network agent centrality scores.  } 
\end{table} \end{center} \vspace{-10mm}

\clearpage
\begin{center}\large{Mean TKO and Base Interaction Network}\end{center}

\begin{center} \begin{table}[!ht] \centering
\scalebox{ 0.83 }{
\rowcolors{2}{tableShade}{}
\begin{tabular}{ccc d{1.6} d{1.6} d{1.6} d{1.6} d{1.6}}
\multicolumn{8}{c}{ \text{Pearson Correlations of Centrality Measures and Mean TKO on Base Network }} \\ \specialrule{.08em}{.00em}{.00em} 
 \text{Disease Type} & \text{Network Type} &\multicolumn{1}{c}{InfectionRate} & \multicolumn{1}{c}{Degree} & \multicolumn{1}{c}{Closeness} & \multicolumn{1}{c}{Betweenness} & \multicolumn{1}{c}{Eigenvector} & \multicolumn{1}{c}{Katz}  \\ \hline

 \text{SIR} & \text{scalefree} & 0.10 & 0.12721 & 0.10583 & 0.09908 & 0.09757 & 0.07997 \\
 \text{SIR} & \text{scalefree} & 0.15 & 0.12514 & 0.09557 & 0.10187 & 0.0977 & 0.08398 \\
 \text{SIR} & \text{scalefree} & 0.20 & 0.13992 & 0.11872 & 0.11503 & 0.11485 & 0.10137 \\ \hline
 \text{SIR} & \text{smallworld} & 0.10 & -0.00492 & 0.03076 & 0.01898 & -0.00766 & -0.01483 \\
 \text{SIR} & \text{smallworld} & 0.15 & 0.04866 & 0.07044 & 0.08208 & 0.08047 & 0.06403 \\
 \text{SIR} & \text{smallworld} & 0.20 & 0.06674 & 0.12695 & 0.10153 & 0.02619 & 0.05665 \\ \hline
 \text{SIS} & \text{scalefree} & 0.10 & 0.13528 & 0.08851 & 0.11109 & 0.09088 & 0.06804 \\
 \text{SIS} & \text{scalefree} & 0.15 & 0.16164 & 0.09822 & 0.127 & 0.10646 & 0.08486 \\
 \text{SIS} & \text{scalefree} & 0.20 & 0.23309 & 0.16747 & 0.19103 & 0.16756 & 0.13834 \\ \hline
 \text{SIS} & \text{smallworld} & 0.10 & 0.02536 & 0.0545 & 0.01982 & -0.0126 & 0.0042 \\
 \text{SIS} & \text{smallworld} & 0.15 & 0.06834 & 0.10208 & 0.11029 & 0.02545 & 0.05613 \\
 \text{SIS} & \text{smallworld} & 0.20 & 0.12315 & 0.17406 & 0.24128 & 0.0522 & 0.12552 \\
 
\specialrule{.08em}{.00em}{.00em} 
\rowcolor{white}
\multicolumn{8}{c}{ \text{  }}\\
\multicolumn{8}{c}{ \text{  }}\\ 
\rowcolor{white}
\multicolumn{8}{c}{ \text{Spearman Correlations of Centrality Measures and Mean TKO on Base Network }} \\ \specialrule{.08em}{.00em}{.00em} 
 \text{Disease Type} & \text{Network Type} &\multicolumn{1}{c}{InfectionRate} & \multicolumn{1}{c}{Degree} & \multicolumn{1}{c}{Closeness} & \multicolumn{1}{c}{Betweenness} & \multicolumn{1}{c}{Eigenvector} & \multicolumn{1}{c}{Katz}  \\ \hline

 \text{SIR} & \text{scalefree} & 0.10 & 0.1905 & 0.09194 & 0.19137 & 0.08058 & 0.04674 \\
 \text{SIR} & \text{scalefree} & 0.15 & 0.20983 & 0.11228 & 0.19027 & 0.1079 & 0.07657 \\
 \text{SIR} & \text{scalefree} & 0.20 & 0.22161 & 0.13071 & 0.21508 & 0.12272 & 0.09371 \\ \hline
 \text{SIR} & \text{smallworld} & 0.10 & 0.00607 & 0.01531 & 0.0249 & 0.02495 & 0.00735 \\
 \text{SIR} & \text{smallworld} & 0.15 & 0.05186 & 0.08794 & 0.08288 & 0.07049 & 0.05397 \\
 \text{SIR} & \text{smallworld} & 0.20 & 0.05635 & 0.15674 & 0.0957 & 0.05611 & 0.05825 \\ \hline
 \text{SIS} & \text{scalefree} & 0.10 & 0.20006 & 0.08366 & 0.2032 & 0.07421 & 0.03927 \\
 \text{SIS} & \text{scalefree} & 0.15 & 0.27654 & 0.10011 & 0.26282 & 0.09862 & 0.05359 \\
 \text{SIS} & \text{scalefree} & 0.20 & 0.32109 & 0.13576 & 0.31428 & 0.12637 & 0.07313 \\ \hline
 \text{SIS} & \text{smallworld} & 0.10 & 0.00761 & 0.02375 & 0.01917 & -0.01435 & -0.01696 \\
 \text{SIS} & \text{smallworld} & 0.15 & 0.06719 & 0.15883 & 0.13379 & 0.0364 & 0.04759 \\
 \text{SIS} & \text{smallworld} & 0.20 & 0.11495 & 0.20274 & 0.21729 & 0.0699 & 0.11563 \\
 
\specialrule{.08em}{.00em}{.00em}  
\rowcolor{white}
\multicolumn{8}{c}{ \text{  }}\\
\multicolumn{8}{c}{ \text{  }}\\
\rowcolor{white}
\multicolumn{8}{c}{ \text{Top Ten Overlap of Centrality Measures and Mean TKO on Base Network }} \\ \specialrule{.08em}{.00em}{.00em} 
 \text{Disease Type} & \text{Network Type} &\multicolumn{1}{c}{InfectionRate} & \multicolumn{1}{c}{Degree} & \multicolumn{1}{c}{Closeness} & \multicolumn{1}{c}{Betweenness} & \multicolumn{1}{c}{Eigenvector} & \multicolumn{1}{c}{Katz}  \\ \hline

 \text{SIR} & \text{scalefree} & 0.10 & 0.012 & 0.008 & 0.012 & 0. & 0. \\
 \text{SIR} & \text{scalefree} & 0.15 & 0.012 & 0.02 & 0. & 0. & 0. \\
 \text{SIR} & \text{scalefree} & 0.20 & 0.02 & 0.004 & 0.016 & 0. & 0. \\ \hline
 \text{SIR} & \text{smallworld} & 0.10 & 0.004 & 0.008 & 0.004 & 0. & 0. \\
 \text{SIR} & \text{smallworld} & 0.15 & 0.008 & 0.024 & 0.008 & 0. & 0. \\
 \text{SIR} & \text{smallworld} & 0.20 & 0.016 & 0.008 & 0.016 & 0. & 0. \\ \hline
 \text{SIS} & \text{scalefree} & 0.10 & 0.012 & 0.008 & 0.016 & 0. & 0. \\
 \text{SIS} & \text{scalefree} & 0.15 & 0.012 & 0.008 & 0.008 & 0. & 0. \\
 \text{SIS} & \text{scalefree} & 0.20 & 0.016 & 0.028 & 0.02 & 0. & 0. \\ \hline
 \text{SIS} & \text{smallworld} & 0.10 & 0.004 & 0.004 & 0.016 & 0. & 0. \\
 \text{SIS} & \text{smallworld} & 0.15 & 0.02 & 0.012 & 0.016 & 0. & 0. \\
 \text{SIS} & \text{smallworld} & 0.20 & 0.012 & 0.016 & 0.032 & 0. & 0. \\

\specialrule{.08em}{.00em}{.00em} 
\end{tabular} } \caption{The Pearson and Spearman correlations as well as the average percent of matching Top Ten agents between the mean proportional TKO score with each of five base network agent centrality scores.} 
\end{table} \end{center} \vspace{-10mm}

\clearpage
\begin{center}\large{Mean TKO and Flattened Interaction Network - Unweighted}\end{center}

\begin{center} \begin{table}[!ht] \centering
\scalebox{ 0.83 }{
\rowcolors{2}{tableShade}{}
\begin{tabular}{ccc d{1.6} d{1.6} d{1.6} d{1.6} d{1.6}}
\multicolumn{8}{c}{ \text{Pearson Correlations of Unweighted Centrality Measures and Mean TKO on Flattened Network }} \\ \specialrule{.08em}{.00em}{.00em} 
 \text{Disease Type} & \text{Network Type} &\multicolumn{1}{c}{InfectionRate} & \multicolumn{1}{c}{Degree} & \multicolumn{1}{c}{Closeness} & \multicolumn{1}{c}{Betweenness} & \multicolumn{1}{c}{Eigenvector} & \multicolumn{1}{c}{Katz}  \\ \hline

 \text{SIR} & \text{scalefree} & 0.10 & 0.12815 & 0.10446 & 0.10138 & 0.06 & -0.0147 \\
 \text{SIR} & \text{scalefree} & 0.15 & 0.12567 & 0.09498 & 0.10378 & 0.06054 & -0.00984 \\
 \text{SIR} & \text{scalefree} & 0.20 & 0.14023 & 0.11871 & 0.1158 & 0.04537 & 0.01406 \\ \hline
 \text{SIR} & \text{smallworld} & 0.10 & -0.00492 & 0.03076 & 0.01898 & -0.0322 & -0.01334 \\
 \text{SIR} & \text{smallworld} & 0.15 & 0.04866 & 0.07044 & 0.08208 & 0.04724 & -0.00565 \\
 \text{SIR} & \text{smallworld} & 0.20 & 0.06674 & 0.12695 & 0.10153 & -0.00722 & 0.01564 \\ \hline
 \text{SIS} & \text{scalefree} & 0.10 & 0.13606 & 0.08709 & 0.11292 & 0.08012 & 0.01291 \\
 \text{SIS} & \text{scalefree} & 0.15 & 0.16208 & 0.09486 & 0.12839 & 0.09913 & 0.01734 \\
 \text{SIS} & \text{scalefree} & 0.20 & 0.23404 & 0.16759 & 0.19278 & 0.10685 & 0.01614 \\ \hline
 \text{SIS} & \text{smallworld} & 0.10 & 0.02536 & 0.0545 & 0.01982 & 0.00807 & 0.00569 \\
 \text{SIS} & \text{smallworld} & 0.15 & 0.06834 & 0.10208 & 0.11029 & 0.0242 & 0.01523 \\
 \text{SIS} & \text{smallworld} & 0.20 & 0.12315 & 0.17406 & 0.24128 & 0.00142 & -0.004 \\
 
\specialrule{.08em}{.00em}{.00em} 
\rowcolor{white}
\multicolumn{8}{c}{ \text{  }}\\
\multicolumn{8}{c}{ \text{  }}\\
\rowcolor{white}
\multicolumn{8}{c}{ \text{Spearman Correlations of Unweighted Centrality Measures and Mean TKO on Flattened Network }} \\ \specialrule{.08em}{.00em}{.00em} 
 \text{Disease Type} & \text{Network Type} &\multicolumn{1}{c}{InfectionRate} & \multicolumn{1}{c}{Degree} & \multicolumn{1}{c}{Closeness} & \multicolumn{1}{c}{Betweenness} & \multicolumn{1}{c}{Eigenvector} & \multicolumn{1}{c}{Katz}  \\ \hline

 \text{SIR} & \text{scalefree} & 0.10 & 0.19049 & 0.09173 & 0.19154 & 0.05292 & -0.0181 \\
 \text{SIR} & \text{scalefree} & 0.15 & 0.20984 & 0.11246 & 0.19295 & 0.07219 & 0.02852 \\
 \text{SIR} & \text{scalefree} & 0.20 & 0.22161 & 0.13124 & 0.2144 & 0.04265 & -0.00284 \\ \hline
 \text{SIR} & \text{smallworld} & 0.10 & 0.00607 & 0.01531 & 0.0249 & -0.04346 & 0.00171 \\
 \text{SIR} & \text{smallworld} & 0.15 & 0.05186 & 0.08794 & 0.08288 & 0.05263 & 0.00793 \\
 \text{SIR} & \text{smallworld} & 0.20 & 0.05635 & 0.15674 & 0.0957 & 0.03002 & 0.00901 \\ \hline
 \text{SIS} & \text{scalefree} & 0.10 & 0.20004 & 0.08303 & 0.2042 & 0.05261 & 0.00635 \\
 \text{SIS} & \text{scalefree} & 0.15 & 0.27653 & 0.09826 & 0.26503 & 0.07858 & 0.04579 \\
 \text{SIS} & \text{scalefree} & 0.20 & 0.32109 & 0.13629 & 0.31345 & 0.05399 & 0.02333 \\ \hline
 \text{SIS} & \text{smallworld} & 0.10 & 0.00761 & 0.02375 & 0.01917 & -0.01027 & -0.00899 \\
 \text{SIS} & \text{smallworld} & 0.15 & 0.06719 & 0.15883 & 0.13379 & 0.06029 & 0.00663 \\
 \text{SIS} & \text{smallworld} & 0.20 & 0.11495 & 0.20274 & 0.21729 & 0.02264 & -0.00918 \\
 
\specialrule{.08em}{.00em}{.00em} 
\rowcolor{white}
\multicolumn{8}{c}{ \text{  }}\\
\multicolumn{8}{c}{ \text{  }}\\
\rowcolor{white}
\multicolumn{8}{c}{ \text{Top Ten Overlap of Unweighted Centrality Measures and Mean TKO on Flattened Network }} \\ \specialrule{.08em}{.00em}{.00em} 
 \text{Disease Type} & \text{Network Type} &\multicolumn{1}{c}{InfectionRate} & \multicolumn{1}{c}{Degree} & \multicolumn{1}{c}{Closeness} & \multicolumn{1}{c}{Betweenness} & \multicolumn{1}{c}{Eigenvector} & \multicolumn{1}{c}{Katz}  \\ \hline
 
 \text{SIR} & \text{scalefree} & 0.10 & 0.008 & 0.004 & 0.008 & 0. & 0. \\
 \text{SIR} & \text{scalefree} & 0.15 & 0.008 & 0.016 & 0.004 & 0. & 0. \\
 \text{SIR} & \text{scalefree} & 0.20 & 0.016 & 0.012 & 0.02 & 0. & 0. \\ \hline
 \text{SIR} & \text{smallworld} & 0.10 & 0.004 & 0.008 & 0.004 & 0. & 0. \\
 \text{SIR} & \text{smallworld} & 0.15 & 0.008 & 0.024 & 0.008 & 0. & 0. \\
 \text{SIR} & \text{smallworld} & 0.20 & 0.016 & 0.008 & 0.016 & 0. & 0. \\ \hline
 \text{SIS} & \text{scalefree} & 0.10 & 0.012 & 0.02 & 0.012 & 0. & 0. \\
 \text{SIS} & \text{scalefree} & 0.15 & 0.016 & 0.008 & 0.004 & 0. & 0. \\
 \text{SIS} & \text{scalefree} & 0.20 & 0.016 & 0.02 & 0.02 & 0. & 0. \\ \hline
 \text{SIS} & \text{smallworld} & 0.10 & 0.004 & 0.004 & 0.016 & 0. & 0. \\
 \text{SIS} & \text{smallworld} & 0.15 & 0.02 & 0.012 & 0.016 & 0. & 0. \\
 \text{SIS} & \text{smallworld} & 0.20 & 0.012 & 0.016 & 0.032 & 0. & 0. \\

\specialrule{.08em}{.00em}{.00em} 
\end{tabular} } \caption{The Pearson and Spearman correlations as well as the average percent of matching Top Ten agents between the mean TKO score with each of five flattened observed interaction network agent centrality scores.  } 
\end{table} \end{center} \vspace{-10mm}

\clearpage
\begin{center}\large{Mean Delta Fraction TKO and Base Interaction Network}\end{center}

\begin{center} \begin{table}[!ht] \centering
\scalebox{ 0.83 }{
\rowcolors{2}{tableShade}{}
\begin{tabular}{ccc d{1.6} d{1.6} d{1.6} d{1.6} d{1.6}}
\multicolumn{8}{c}{ \text{Pearson Correlations of Centrality Measures and Mean Delta TKO on Base Network }} \\ \specialrule{.08em}{.00em}{.00em} 
 \text{Disease Type} & \text{Network Type} &\multicolumn{1}{c}{InfectionRate} & \multicolumn{1}{c}{Degree} & \multicolumn{1}{c}{Closeness} & \multicolumn{1}{c}{Betweenness} & \multicolumn{1}{c}{Eigenvector} & \multicolumn{1}{c}{Katz}  \\ \hline

 \text{SIR} & \text{scalefree} & 0.10 & 0.11126 & 0.09799 & 0.08769 & 0.0877 & 0.07232 \\
 \text{SIR} & \text{scalefree} & 0.15 & 0.14889 & 0.11732 & 0.12976 & 0.12374 & 0.11078 \\
 \text{SIR} & \text{scalefree} & 0.20 & 0.16805 & 0.14606 & 0.13956 & 0.14173 & 0.12793 \\ \hline
 \text{SIR} & \text{smallworld} & 0.10 & -0.00009 & 0.04849 & 0.0285 & -0.01346 & -0.015 \\
 \text{SIR} & \text{smallworld} & 0.15 & 0.07554 & 0.15629 & 0.11958 & 0.08021 & 0.0855 \\
 \text{SIR} & \text{smallworld} & 0.20 & 0.10399 & 0.17922 & 0.14133 & 0.03744 & 0.0903 \\ \hline
 \text{SIS} & \text{scalefree} & 0.10 & 0.1261 & 0.08495 & 0.10497 & 0.08869 & 0.06882 \\
 \text{SIS} & \text{scalefree} & 0.15 & 0.18292 & 0.11692 & 0.14573 & 0.12508 & 0.10177 \\
 \text{SIS} & \text{scalefree} & 0.20 & 0.24463 & 0.17563 & 0.20131 & 0.17664 & 0.14665 \\ \hline
 \text{SIS} & \text{smallworld} & 0.10 & 0.02352 & 0.08397 & 0.02964 & -0.01187 & 0.00138 \\
 \text{SIS} & \text{smallworld} & 0.15 & 0.09169 & 0.22063 & 0.15242 & 0.04315 & 0.07855 \\
 \text{SIS} & \text{smallworld} & 0.20 & 0.14818 & 0.2428 & 0.2825 & 0.0684 & 0.15092 \\
 
\specialrule{.08em}{.00em}{.00em} 
\rowcolor{white}
\multicolumn{8}{c}{ \text{  }}\\
\multicolumn{8}{c}{ \text{  }}\\ 
\rowcolor{white}
\multicolumn{8}{c}{ \text{Spearman Correlations of Centrality Measures and Mean Delta TKO on Base Network }} \\ \specialrule{.08em}{.00em}{.00em} 
 \text{Disease Type} & \text{Network Type} &\multicolumn{1}{c}{InfectionRate} & \multicolumn{1}{c}{Degree} & \multicolumn{1}{c}{Closeness} & \multicolumn{1}{c}{Betweenness} & \multicolumn{1}{c}{Eigenvector} & \multicolumn{1}{c}{Katz}  \\ \hline

 \text{SIR} & \text{scalefree} & 0.10 & 0.1898 & 0.09614 & 0.1907 & 0.08385 & 0.04946 \\
 \text{SIR} & \text{scalefree} & 0.15 & 0.21847 & 0.12032 & 0.20791 & 0.11745 & 0.08826 \\
 \text{SIR} & \text{scalefree} & 0.20 & 0.22746 & 0.13761 & 0.21875 & 0.1287 & 0.09995 \\ \hline
 \text{SIR} & \text{smallworld} & 0.10 & 0.01018 & 0.03939 & 0.03931 & 0.02556 & 0.00609 \\
 \text{SIR} & \text{smallworld} & 0.15 & 0.0516 & 0.14602 & 0.08077 & 0.0788 & 0.0599 \\
 \text{SIR} & \text{smallworld} & 0.20 & 0.0699 & 0.21156 & 0.12171 & 0.05664 & 0.06476 \\ \hline
 \text{SIS} & \text{scalefree} & 0.10 & 0.19765 & 0.08699 & 0.2017 & 0.07697 & 0.04237 \\
 \text{SIS} & \text{scalefree} & 0.15 & 0.29223 & 0.10894 & 0.28278 & 0.10884 & 0.06348 \\
 \text{SIS} & \text{scalefree} & 0.20 & 0.32746 & 0.13763 & 0.32083 & 0.12769 & 0.07348 \\ \hline
 \text{SIS} & \text{smallworld} & 0.10 & 0.00888 & 0.05328 & 0.03496 & -0.02485 & -0.02329 \\
 \text{SIS} & \text{smallworld} & 0.15 & 0.07207 & 0.24837 & 0.14039 & 0.05486 & 0.05916 \\
 \text{SIS} & \text{smallworld} & 0.20 & 0.11845 & 0.25568 & 0.22694 & 0.07612 & 0.11974 \\
 
\specialrule{.08em}{.00em}{.00em}  
\rowcolor{white}
\multicolumn{8}{c}{ \text{  }}\\
\multicolumn{8}{c}{ \text{  }}\\
\rowcolor{white}
\multicolumn{8}{c}{ \text{Top Ten Overlap of Centrality Measures and Mean Delta TKO on Base Network }} \\ \specialrule{.08em}{.00em}{.00em} 
 \text{Disease Type} & \text{Network Type} &\multicolumn{1}{c}{InfectionRate} & \multicolumn{1}{c}{Degree} & \multicolumn{1}{c}{Closeness} & \multicolumn{1}{c}{Betweenness} & \multicolumn{1}{c}{Eigenvector} & \multicolumn{1}{c}{Katz}  \\ \hline

 \text{SIR} & \text{scalefree} & 0.10 & 0.016 & 0.012 & 0.012 & 0. & 0. \\
 \text{SIR} & \text{scalefree} & 0.15 & 0.028 & 0.016 & 0.032 & 0. & 0. \\
 \text{SIR} & \text{scalefree} & 0.20 & 0.024 & 0.012 & 0.02 & 0. & 0. \\ \hline
 \text{SIR} & \text{smallworld} & 0.10 & 0.008 & 0. & 0. & 0. & 0. \\
 \text{SIR} & \text{smallworld} & 0.15 & 0.012 & 0.02 & 0.008 & 0. & 0. \\
 \text{SIR} & \text{smallworld} & 0.20 & 0.02 & 0.02 & 0.012 & 0. & 0. \\ \hline
 \text{SIS} & \text{scalefree} & 0.10 & 0.012 & 0.024 & 0.016 & 0. & 0. \\
 \text{SIS} & \text{scalefree} & 0.15 & 0.024 & 0.004 & 0.012 & 0. & 0. \\
 \text{SIS} & \text{scalefree} & 0.20 & 0.024 & 0.044 & 0.028 & 0. & 0. \\ \hline
 \text{SIS} & \text{smallworld} & 0.10 & 0.012 & 0.008 & 0.02 & 0. & 0. \\
 \text{SIS} & \text{smallworld} & 0.15 & 0.012 & 0.02 & 0.032 & 0. & 0. \\
 \text{SIS} & \text{smallworld} & 0.20 & 0.012 & 0.012 & 0.028 & 0. & 0. \\

\specialrule{.08em}{.00em}{.00em} 
\end{tabular} } \caption{The Pearson and Spearman correlations as well as the average percent of matching Top Ten agents between the mean change in fractional TKO score with each of five base network agent centrality scores.} 
\end{table} \end{center} \vspace{-10mm}

\clearpage
\begin{center}\large{Mean Delta Fraction TKO and Flattened Interaction Network - Unweighted}\end{center}

\begin{center} \begin{table}[!ht] \centering
\scalebox{ 0.83 }{
\rowcolors{2}{tableShade}{}
\begin{tabular}{ccc d{1.6} d{1.6} d{1.6} d{1.6} d{1.6}}
\multicolumn{8}{c}{ \text{Pearson Correlations of Unweighted Centrality Measures and Mean Delta TKO on Flattened Network }} \\ \specialrule{.08em}{.00em}{.00em} 
 \text{Disease Type} & \text{Network Type} &\multicolumn{1}{c}{InfectionRate} & \multicolumn{1}{c}{Degree} & \multicolumn{1}{c}{Closeness} & \multicolumn{1}{c}{Betweenness} & \multicolumn{1}{c}{Eigenvector} & \multicolumn{1}{c}{Katz}  \\ \hline

 \text{SIR} & \text{scalefree} & 0.10 & 0.11209 & 0.09676 & 0.08987 & 0.04575 & -0.00955 \\
 \text{SIR} & \text{scalefree} & 0.15 & 0.14898 & 0.11589 & 0.13081 & 0.07341 & 0.00383 \\
 \text{SIR} & \text{scalefree} & 0.20 & 0.16834 & 0.14605 & 0.14042 & 0.05829 & 0.01177 \\ \hline
 \text{SIR} & \text{smallworld} & 0.10 & -0.00009 & 0.04849 & 0.0285 & -0.03495 & -0.00892 \\
 \text{SIR} & \text{smallworld} & 0.15 & 0.07554 & 0.15629 & 0.11958 & 0.06106 & -0.00153 \\
 \text{SIR} & \text{smallworld} & 0.20 & 0.10399 & 0.17922 & 0.14133 & -0.00044 & 0.00307 \\ \hline
 \text{SIS} & \text{scalefree} & 0.10 & 0.12659 & 0.08274 & 0.10637 & 0.07632 & 0.01533 \\
 \text{SIS} & \text{scalefree} & 0.15 & 0.18318 & 0.11273 & 0.14702 & 0.1164 & 0.02053 \\
 \text{SIS} & \text{scalefree} & 0.20 & 0.24569 & 0.17587 & 0.20314 & 0.11365 & 0.01806 \\ \hline
 \text{SIS} & \text{smallworld} & 0.10 & 0.02352 & 0.08397 & 0.02964 & -0.00576 & 0.01179 \\
 \text{SIS} & \text{smallworld} & 0.15 & 0.09169 & 0.22063 & 0.15242 & 0.04087 & 0.01275 \\
 \text{SIS} & \text{smallworld} & 0.20 & 0.14818 & 0.2428 & 0.2825 & -0.00623 & 0.00016 \\
 
\specialrule{.08em}{.00em}{.00em} 
\rowcolor{white}
\multicolumn{8}{c}{ \text{  }}\\
\multicolumn{8}{c}{ \text{  }}\\
\rowcolor{white}
\multicolumn{8}{c}{ \text{Spearman Correlations of Unweighted Centrality Measures and Mean Delta TKO on Flattened Network }} \\ \specialrule{.08em}{.00em}{.00em} 
 \text{Disease Type} & \text{Network Type} &\multicolumn{1}{c}{InfectionRate} & \multicolumn{1}{c}{Degree} & \multicolumn{1}{c}{Closeness} & \multicolumn{1}{c}{Betweenness} & \multicolumn{1}{c}{Eigenvector} & \multicolumn{1}{c}{Katz}  \\ \hline

 \text{SIR} & \text{scalefree} & 0.10 & 0.18978 & 0.09581 & 0.19107 & 0.05107 & -0.02437 \\
 \text{SIR} & \text{scalefree} & 0.15 & 0.21847 & 0.12017 & 0.21048 & 0.06814 & 0.02238 \\
 \text{SIR} & \text{scalefree} & 0.20 & 0.22745 & 0.13857 & 0.21806 & 0.04818 & 0.00276 \\ \hline
 \text{SIR} & \text{smallworld} & 0.10 & 0.01018 & 0.03939 & 0.03931 & -0.05202 & 0.00721 \\
 \text{SIR} & \text{smallworld} & 0.15 & 0.0516 & 0.14602 & 0.08077 & 0.09211 & 0.01342 \\
 \text{SIR} & \text{smallworld} & 0.20 & 0.0699 & 0.21156 & 0.12171 & 0.03232 & 0.00737 \\ \hline
 \text{SIS} & \text{scalefree} & 0.10 & 0.19763 & 0.08611 & 0.20295 & 0.05174 & 0.01168 \\
 \text{SIS} & \text{scalefree} & 0.15 & 0.29223 & 0.10699 & 0.28511 & 0.08752 & 0.04597 \\
 \text{SIS} & \text{scalefree} & 0.20 & 0.32746 & 0.13833 & 0.31985 & 0.05933 & 0.02361 \\ \hline
 \text{SIS} & \text{smallworld} & 0.10 & 0.00888 & 0.05328 & 0.03496 & -0.00993 & -0.00835 \\
 \text{SIS} & \text{smallworld} & 0.15 & 0.07207 & 0.24837 & 0.14039 & 0.11152 & 0.00933 \\
 \text{SIS} & \text{smallworld} & 0.20 & 0.11845 & 0.25568 & 0.22694 & 0.01745 & -0.01791 \\
 
\specialrule{.08em}{.00em}{.00em} 
\rowcolor{white}
\multicolumn{8}{c}{ \text{  }}\\
\multicolumn{8}{c}{ \text{  }}\\
\rowcolor{white}
\multicolumn{8}{c}{ \text{Top Ten Overlap of Unweighted Centrality Measures and Mean Delta TKO on Flattened Network }} \\ \specialrule{.08em}{.00em}{.00em} 
 \text{Disease Type} & \text{Network Type} &\multicolumn{1}{c}{InfectionRate} & \multicolumn{1}{c}{Degree} & \multicolumn{1}{c}{Closeness} & \multicolumn{1}{c}{Betweenness} & \multicolumn{1}{c}{Eigenvector} & \multicolumn{1}{c}{Katz}  \\ \hline
 
 \text{SIR} & \text{scalefree} & 0.10 & 0.016 & 0.02 & 0.012 & 0. & 0. \\
 \text{SIR} & \text{scalefree} & 0.15 & 0.024 & 0.008 & 0.024 & 0. & 0. \\
 \text{SIR} & \text{scalefree} & 0.20 & 0.028 & 0.024 & 0.016 & 0. & 0. \\ \hline
 \text{SIR} & \text{smallworld} & 0.10 & 0.008 & 0. & 0. & 0. & 0. \\
 \text{SIR} & \text{smallworld} & 0.15 & 0.012 & 0.02 & 0.008 & 0. & 0. \\
 \text{SIR} & \text{smallworld} & 0.20 & 0.02 & 0.02 & 0.012 & 0. & 0. \\ \hline
 \text{SIS} & \text{scalefree} & 0.10 & 0.012 & 0.016 & 0.012 & 0. & 0. \\
 \text{SIS} & \text{scalefree} & 0.15 & 0.02 & 0.004 & 0.012 & 0. & 0. \\
 \text{SIS} & \text{scalefree} & 0.20 & 0.028 & 0.024 & 0.02 & 0. & 0. \\ \hline
 \text{SIS} & \text{smallworld} & 0.10 & 0.012 & 0.008 & 0.02 & 0. & 0. \\
 \text{SIS} & \text{smallworld} & 0.15 & 0.012 & 0.02 & 0.032 & 0. & 0. \\
 \text{SIS} & \text{smallworld} & 0.20 & 0.012 & 0.012 & 0.028 & 0. & 0. \\

\specialrule{.08em}{.00em}{.00em} 
\end{tabular} } \caption{The Pearson and Spearman correlations as well as the average percent of matching Top Ten agents between the mean change in fractional TKO score with each of five flattened observed interaction network agent centrality scores.  } 
\end{table} \end{center} \vspace{-10mm}

\end{document}